\documentclass[10pt]{article} % onecolumn (standard format)

\usepackage[utf8]{inputenc}
\usepackage{amsmath}
\usepackage{amsfonts}
\usepackage{amssymb}
\usepackage{mathtools}
\usepackage{listings}
\usepackage{mathrsfs}
\usepackage{times}
\usepackage{float}
\usepackage{color}
\usepackage{setspace}
\usepackage{color,soul}

\usepackage{amsmath,amsfonts,amsthm,eucal}
\usepackage{enumerate}
\usepackage[letterpaper,margin=3cm]{geometry}
\usepackage{float}
\usepackage[title]{appendix}
\usepackage{color}
\usepackage{tabularx}
\usepackage{siunitx}
\usepackage[tableposition=below]{caption}
\usepackage[
pdfstartview=XYZ,
bookmarks=true,
colorlinks=true,
linkcolor=blue,
urlcolor=blue,
citecolor=blue,
pdftex,
bookmarks=true,
linktocpage=true,   % makes the page number as hyperlink in table of content
hyperindex=true
]{hyperref}

\usepackage{lipsum}
\usepackage[FIGBOTCAP,TABTOPCAP,bf,tight]{subfigure}

\usepackage{natbib}
\usepackage[pdftex]{graphicx}
\usepackage{epstopdf}
\usepackage{booktabs} 
\usepackage{tikz,mathpazo}
\usetikzlibrary{shapes.geometric, arrows}
\usepackage{caption}

\newcommand{\tensor}[1]{\ensuremath{\boldsymbol{#1}}}

\DeclareMathOperator{\grad}{\nabla}

\DeclareMathOperator{\diver}{\nabla\cdot}

\theoremstyle{remark}
\newtheorem{rmk}{Remark}
\renewcommand{\vec}[1]{\ensuremath{\boldsymbol{#1}}}

\usepackage{algorithm}
\usepackage{algorithmicx}
\usepackage[noend]{algpseudocode}

\theoremstyle{definition}

\usepackage{longtable}
\usepackage{adjustbox}
\usepackage{framed}

\title{A phase field model for cohesive fracture in micropolar continua} 

\begin{document}

\author{Hyoung Suk Suh\thanks{Department of Civil Engineering and Engineering Mechanics, 
 Columbia University, 
 New York, NY 10027.     \textit{h.suh@columbia.edu}  }       \and
        WaiChing Sun\thanks{Department of Civil Engineering and Engineering Mechanics, 
 Columbia University, 
 New York, NY 10027.
  \textit{wsun@columbia.edu}  (corresponding author)      %     \\
}
\and
Devin T. O'Connor\thanks{Sea Ice Research Center, Cold Regions Research and Engineering Laboratory, Hanover, NH 03755.
  \textit{devin.t.oconnor@erdc.dren.mil}}
}

\maketitle

\begin{abstract}
While crack nucleation and propagation in the brittle or quasi-brittle regime can be predicted via variational or material-force-based phase field fracture models, these models often assume that the underlying elastic response of the material is non-polar and yet a length scale parameter must be introduced to enable the sharp cracks represented by a regularized implicit function. 
However, many materials with internal microstructures that contain surface tension, micro-cracks, micro-fracture, inclusion, cavity or those of particulate nature often exhibit size-dependent behaviors in both the path-independent and path-dependent regimes. 
This paper is intended to introduce a unified treatment that captures the size effect of the materials in both elastic and damaged states. 
By introducing a cohesive micropolar phase field fracture theory, along with the computational model and validation exercises, we explore the interacting size-dependent elastic deformation and fracture mechanisms exhibits in materials of complex microstructures. 
To achieve this goal, we introduce the distinctive degradation functions of the force-stress-strain and couple-stress-micro-rotation energy-conjugated pairs for a given regularization profile such that the macroscopic size-dependent responses of the micropolar continua is insensitive to the length scale parameter of the regularized interface. 
Then, we apply the variational principle to derive governing equations from the micropolar stored energy and dissipative functionals. 
Numerical examples are introduced to demonstrate the proper way to identify material parameters and the capacity of the new formulation to simulate complex crack patterns in the quasi-static regime. 
\end{abstract}

\section{Introduction}
\label{intro}
The size effect and the corresponding length scale parameter associated with the phase field fracture model for brittle or quasi-brittle materials have been a subject of intensive research in recent years \citep{francfort1998revisiting, Bourdin2008, de2016gradient, wang2017unified, aldakheel2018phase, wu2018length, geelen2018optimization, Bryant2018, choo2018cracking, qinami2019circumventing, noii2020adaptive}.
Due to the fact that the phase field approach employ regularized (smoothed) implicit function to represent sharp interface, the physical interpretation of the length scale parameter (and in some cases, the lack thereof) has become a hotly debated topic among the computational fracture mechanics community. 
The lack of consensus on the definition of length scale has also been sometimes perceived as a weakness, especially when compared with the embedded discontinuity approaches such as XFEM or assumed strain models \citep{moes1999finite, armero2008new, wang2018multiscale, wang2019meta, wang2019updated}.

The early attempt to justify the introduction of the length scale parameter for phase field fracture models can be tracked back to the first variational fracture model in \citet{francfort1998revisiting} where the variational fracture model is expected to exhibit $\Gamma-$convergence and therefore may converge to the sharp interface model as the mesh size and the length scale parameter approaches zero. 
While this line of work (e.g., \citet{Bourdin2008} and \citet{may2015numerical}) provides a theoretical justification, in practice, the length scale parameter must still be sufficiently large compared to the mesh size in order to solve the phase field governing equation. 
This could be problematic if the simulations are designed for boundary value problems at the large scales (e.g., hydraulic fracture, faulting of geological formation) where the small mesh size, even concentrated at a local regions, become impractical.

One strategy to overcome this issue is to decouple or eliminate the effect of the length scale parameters on the constitutive responses such that a relatively large length scale parameter can be used for numerical purposes without compromising the accuracy of the constitutive responses. 
To derive a phase field fracture model that exhibits macroscopic responses independent or at least not sensitive to the length scale parameters, \citet{wu2018length} and \citet{Geelen2019} both introduce new crack surface density functionals and the corresponding degradation function derived from cohesive zone models such that the underlying traction-separation law is independent of the length scale parameter. 
Their simulations have shown that the resultant fracture patterns and the macroscopic constitutive responses are both insensitive to the length scale parameters in the sub-critical regime.

Another strategy to circumvent the length scale issue is to determine the underlying relationship among the length scale parameters and other material parameters (e.g., Young's modulus, tensile strength) that can be obtained from a specific set of experimental tests \citep{Nguyen2016, pham2017experimental, choo2018coupled}. 
However, these previous works also show that the analytical expression of the length scale parameter may vary in according to the chosen inverse problems and hence the length scale parameter is likely only valid for backward calibration for a specific problem but cannot be used for general-purposed forward predictions. 
Furthermore, identifying the correct length scale parameter for a given inverse problem does not imply that such a length scale parameter is sufficiently large to ensure the solvability of the discretized governing equation(s).

Nevertheless, \textit{one key aspect that is often overlooked in the phase field fracture modeling is that materials with internal structures that enables size effects on damage and fracture may likely exhibit size effect in the elastic regimes}. 
Examples of these materials include concrete, composite, particulate materials as well as some metamaterials \citep{dietsche1993micropolar, bazant1997fracture, trovalusci2014particulate, wang2016identifying, aldakheel2020microscale}. 
Since these materials exhibit large internal length scales compared to the length scale of damage or fracture, suitable size effect must be carefully incorporated in \textit{both} the elastic and path-dependent regimes to capture the size-dependence properly.
This paper is the first attempt to formulate a new cohesive micropolar phase field fracture theory that leads to a physically justified/identifiable size-dependent effect for both the path-independent elastic responses and the path-dependent damage and fracture in a higher-order continuum undergoing infinitesimal deformation. 
By extending the length-scale-parameter insensitive formulation that approximates cohesive-type of response to the micropolar materials, we introduce a third strategy where one may employ sufficiently large phase field length scale parameters to address the numerical needs without comprising the correct size effect that should exhibit in the numerical simulations. 
In order words, the resultant model is the best of both worlds, one that benefits from the physical justification of having a consistent size effect in both the elastic and damage regimes and yet retains the convenience of the approach in \citet{wu2018length}, \citet{Geelen2019}, and \citet{wu2020phase}.

The rest of the paper is organized as follows. 
We first briefly summarize the theory of micropolar elasticity (Section \ref{micropolar_elasticity}), and introduce the strain energy split approach that enables one to explore the effects of the partitioned energy densities. 
We then extend the regularized length-scale-insensitive phase field formulation to the micropolar material models such that the length scale parameter for the phase field is insensitive to the macroscopic responses. 
This treatment then enables us to replicate the size effect characterized by the higher-order material parameters (e.g., bending and torsion stiffnesses) that can be experimentally sought. 
Furthermore, by enforcing the macroscopic responses not sensitive to the phase field length scale parameter, it enables us to conduct simulations in a spatial domain without the size constraint imposed by the ratio between the phase field length scale and mesh. 
For completeness, the details of the finite element discretization and operator-split solution scheme are discussed. 
Numerical examples are given to verify the implementation, provide evidences on how micropolarity affects the macroscopic behaviors for quasi-brittle materials, and showcase the applicability of the proposed models.

As for notations and symbols, bold-faced and blackboard bold-faced letters denote tensors (including vectors which are rank-one tensors); 
the symbol '$\cdot$' denotes a single contraction of adjacent indices of two tensors 
(e.g.,\ $\vec{a} \cdot \vec{b} = a_{i}b_{i}$ or $\tensor{c} \cdot \tensor{d} = c_{ij}d_{jk}$); 
the symbol `:' denotes a double contraction of adjacent indices of tensor of rank two or higher
(e.g.,\ $\mathbb{C} : \vec{\varepsilon}$ = $C_{ijkl} \varepsilon_{kl}$); 
the symbol `$\otimes$' denotes a juxtaposition of two vectors 
(e.g.,\ $\vec{a} \otimes \vec{b} = a_{i}b_{j}$)
or two symmetric second order tensors 
[e.g.,\ $(\tensor{\alpha} \otimes \tensor{\beta})_{ijkl} = \alpha_{ij}\beta_{kl}$]. 
We also define identity tensors: $\tensor{I} = \delta_{ij}$, $\mathbb{I} = \delta_{ik}\delta_{jl}$, and $\bar{\mathbb{I}} = \delta_{il}\delta_{jk}$, where $\delta_{ij}$ is the Kronecker delta.
As for sign conventions, unless specified, the directions of the tensile stress and dilative pressure are considered as positive.

\section{Theory of micropolar elasticity}
\label{micropolar_elasticity}
In this section, we briefly summarize the kinematic and constitutive relations of an isotropic micropolar elastic materials undergoing infinitesimal deformation. 
In this case,  the kinematics of micropolar materials  is characterized by both the displacement field and the micro-rotations. The resultant strain tensor is no longer symmetric due to the higher-order kinematics. 
We thus decompose the strain tensor into symmetric and skew-symmetric parts, which consequently enables us to split the stored energy density into three different parts. 
This energy split approach opens the door for us to explore the effects of distinct degradation of partitioned energy conjugated pairs, which will be discussed later in this study.

\subsection{Kinematics}
\label{micropolar_kinematics}
\begin{figure}[h]
\centering
\includegraphics[height=0.20\textwidth]{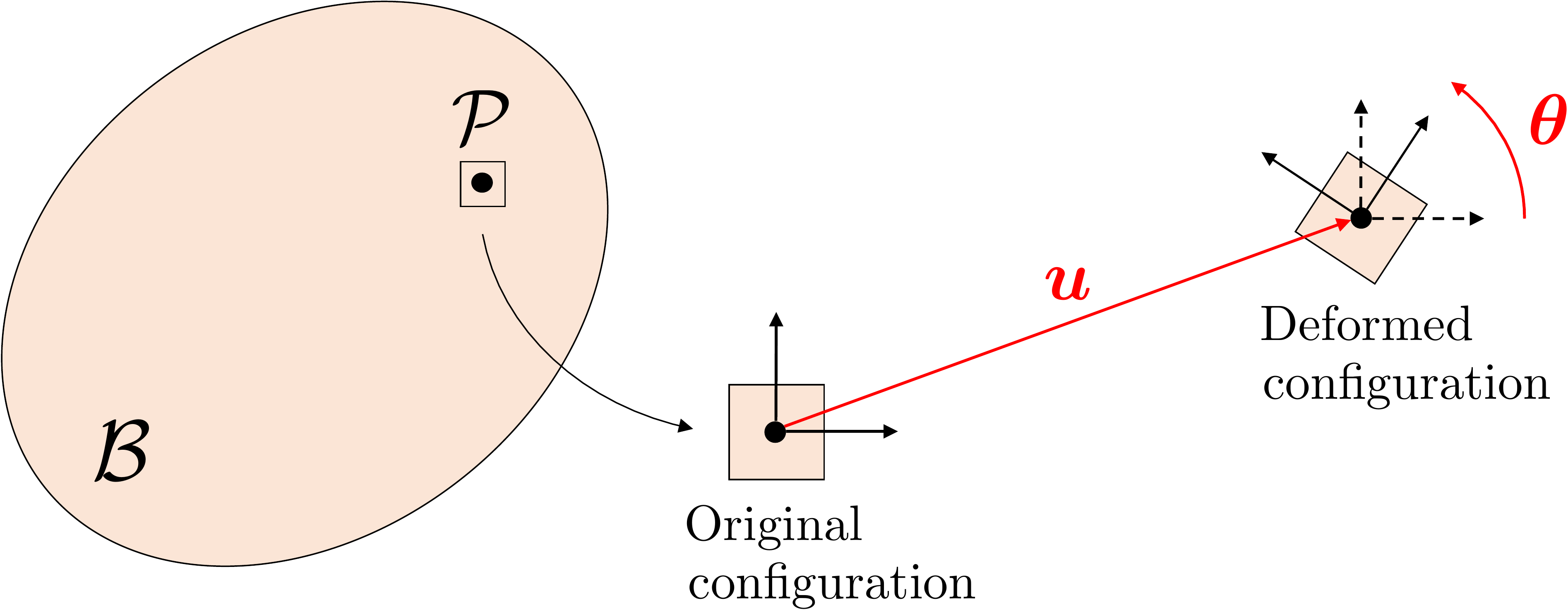}
\caption{Kinematics of a micropolar continuum.}
\label{fig:micropolar_kinematics}
\end{figure}

Let us consider a micropolar elastic body $\mathcal{B} \subset \mathbb{R}^3$ with material points $\mathcal{P}$ identified by the position vectors $\vec{x} \in \mathcal{B}$ that undergoes infinitesimal deformation. 
As illustrated in Fig.~\ref{fig:micropolar_kinematics}, unlike the classical non-polar (Boltzmann) approach, each material point experiences micro-rotation $\vec{\theta}(\vec{x}, t)$, in addition to the translational displacement $\vec{u}(\vec{x}, t)$ at time $t$. 
The micro-rotation represents the local rotation of the material point $\vec{x}$, which is
independent of the displacement field. 
Consequently, the rotational part of the polar decomposition of the displacement gradient (i.e., macro-rotation) is also independent of the micro-rotation. 
The micropolar strain $\bar{\tensor{\varepsilon}}$ and micro-curvature $\bar{\tensor{\kappa}}$ can be defined as follows \citep{eringen1966linear, sachio1984finite, ehlers1998theoretical, eringen2012microcontinuum, atroshchenko2015fundamental, grbvcic2018variational}:
\begin{align}
\label{eq:micropolar_strain}
& \bar{\tensor{\varepsilon}}
= \grad{\vec{u}}^{\text{T}} - \overset{3}{\tensor{E}} \cdot \vec{\theta}, \\
\label{eq:micro-curvature}
& \bar{\tensor{\kappa}} = \grad{\vec{\theta}},
\end{align}
where $\overset{3}{\tensor{E}} = \overset{3}{E}_{ijk}$ is the Levi-Civita permutation tensor. 
The definition of micropolar strain in Eq.~\eqref{eq:micropolar_strain} implies that the normal strains (i.e., diagonal entries of the micropolar strain tensor) that contributes to the stretching are equivalent to those in the classical approach, whereas the shear strains (i.e., off-diagonal entries of the micropolar strain tensor) are dependent on the micro-rotation. 
Since the micropolar strain tensor is non-symmetric, we therefore decompose the micropolar strain tensor into symmetric ($\bar{\tensor{\varepsilon}}^{\text{sym}}$) and skew-symmetric parts ($\bar{\tensor{\varepsilon}}^{\text{skew}}$), i.e.,
\begin{equation}
\label{eq:strain_split1}
\bar{\tensor{\varepsilon}} = 
\underbrace{
\frac{1}{2} ( \grad{\vec{u}} + \grad{\vec{u}}^{\text{T}} ) 
}_{:= \bar{\tensor{\varepsilon}}^{\text{sym}}}
+ 
\underbrace{
\frac{1}{2} ( \grad{\vec{u}}^{\text{T}} - \grad{\vec{u}} ) - \overset{3}{\tensor{E}} \cdot \vec{\theta}
}_{:= \bar{\tensor{\varepsilon}}^{\text{skew}}}.
\end{equation}
Notice that $\bar{\tensor{\varepsilon}}^{\text{sym}}$ is equivalent to the Boltzmann strain tensor in classical non-polar approach. 

\subsection{Constitutive model and strain energy split}
\label{constitutive_and_energy_split}
To ensure stable elastic responses, the micropolar strain energy must fulfill strong ellipticity condition. 
To fulfill this requirement, we consider a micropolar strain energy density $\psi_e ( \bar{\tensor{\varepsilon}}, \bar{\tensor{\kappa}} )$ that takes a quadratic form:
\begin{equation}
\label{eq:micropolar_energy}
\psi_e ( \bar{\tensor{\varepsilon}}, \bar{\tensor{\kappa}} )
= \frac{1}{2} \bar{\tensor{\varepsilon}} : \mathbb{C} : \bar{\tensor{\varepsilon}} 
+ \frac{1}{2} \bar{\tensor{\kappa}} : \mathbb{D} : \bar{\tensor{\kappa}}.
\end{equation}
Here, $\mathbb{C}$ and $\mathbb{D}$ are constitutive moduli that possess the major symmetry (i.e., $\mathbb{C} = C_{ijkl} = C_{klij}$, and $\mathbb{D} = D_{ijkl} = D_{klij}$):
\begin{equation}
\label{eq:micropolar_moduli}
\mathbb{C} 
= 
\lambda \left( \tensor{I} \otimes \tensor{I} \right)
+
\left( \mu + \kappa \right) \mathbb{I}
+
\mu \bar{\mathbb{I}}
\: \: ; \: \:
\mathbb{D} = 
\alpha \left( \tensor{I} \otimes \tensor{I} \right)
+
\beta \bar{\mathbb{I}}
+
\gamma \mathbb{I},
\end{equation}
where $\lambda$, $\mu$, $\kappa$, $\alpha$, $\beta$, and $\gamma$ are the material constants. 
The strong ellipticity of the strain energy density defined in Eq. \eqref{eq:micropolar_energy} implies that the following inequalities must be hold \citep{diegele2004linear, li2009fracture, eringen2012microcontinuum}: 
\begin{equation}
\begin{aligned}
\label{eq:materialconstant_conditions}
& 3 \lambda + 2 \mu + \kappa \geq 0 \: ; && 2 \mu + \kappa \geq 0 \: ; && \kappa \geq 0 \: ;
\\
& 3\alpha + \beta + \gamma \geq 0 \: ; && \gamma + \beta \geq 0 \: ; && \gamma - \beta \geq 0.
\end{aligned}
\end{equation}
These material constants, including the size-dependent ones are related to the following material parameters that have been individually identified via experiments \citep{yavari2002fractal, atroshchenko2015fundamental,lakes2015bending}:
\begin{equation}
\label{eq:engineering_material_constants}
\begin{dcases}
E = \frac{\left( 2 \mu + \kappa \right) \left( 3 \lambda + 2 \mu + \kappa \right)}{2 \lambda + 2 \mu + \kappa} & \text{Young's modulus}, \\
G = \frac{2 \mu + \kappa}{2} & \text{shear modulus}, \\
\nu = \frac{\lambda}{2 \lambda + 2 \mu + \kappa} & \text{Poisson's ratio}, \\
l_t = \sqrt{\frac{\beta + \gamma}{2 \mu + \kappa}} & \text{characteristic length in torsion}, \\
l_b = \sqrt{\frac{\gamma}{2 \left( 2 \mu + \kappa \right)}} & \text{characteristic length in bending}, \\
\chi = \frac{\beta + \gamma}{\alpha + \beta + \gamma} & \text{polar ratio}, \\
N = \sqrt{\frac{\kappa}{2 \left( \mu + \kappa \right)}} & \text{coupling number}, N \in [0, 1].
\end{dcases}
\end{equation}
The relations in  Eq.~\eqref{eq:engineering_material_constants} indicates that the size-dependence of the elasticity responses are related to the higher-order kinematics and kinetic. 
Although identification of the material parameters that characterize the size effect remains challenging as demonstrated in \citet{bigoni2007analytical} and \citet{neff2010stable}, these micropolar material parameters are obtainable through well-documented inverse problems or analytical solutions, at least for a subset of micropolar materials such as porous media \citep{bigoni2007analytical}.  
This unambiguity is helpful for practical purposes.

In Eq.~\eqref{eq:engineering_material_constants}, the characteristic lengths $l_t$ and $l_b$ imply the nonlocal nature of micropolar material by quantifying the range of couple stress through their relationship to the micro-curvature. 
The coupling number $N$, on the other hand, quantifies the level of shear stress asymmetry that represents the degree of micropolarity of the material, e.g., $N = 0$ corresponds to the classical elasticity while $N=1$ corresponds to the couple-stress theory \citep{mindlin1962effects, mindlin1963influence,mcgregor2014coupling}. 
In the remainder of this paper, unless specified, we set $N = 0.5$ for micropolar continuum simulations.

Since this study aims to develop a framework that explores the interaction between size-dependent micropolar elasticity and fracture mechanisms, we split the strain energy density into three different parts [based on the decomposition of the micropolar strain, cf. Eq.~\eqref{eq:strain_split1}]: (1) the Boltzmann part $\psi_e^B  ( \bar{\tensor{\varepsilon}}^{\text{sym}} )$; (2) the micro-continuum coupling part $\psi_e^C ( \bar{\tensor{\varepsilon}}^{\text{skew}} )$; and (3) the pure micro-rotational part $\psi_e^R ( \bar{\tensor{\kappa}} )$, i.e.,
\begin{equation}
\label{eq:energy_split}
\psi_e ( \bar{\tensor{\varepsilon}}, \bar{\tensor{\kappa}} )
= \psi_e^B ( \bar{\tensor{\varepsilon}}^{\text{sym}} )
+ \psi_e^C ( \bar{\tensor{\varepsilon}}^{\text{skew}} )
+ \psi_e^R ( \bar{\tensor{\kappa}} ),
\end{equation} 
where the partitioned strain energy densities can be written as:
\begin{align}
\label{eq:energy_parts_B}
& \psi_e^B ( \bar{\tensor{\varepsilon}}^{\text{sym}} ) =
\frac{1}{2} 
\left[ \lambda ( \bar{\tensor{\varepsilon}}^{\text{sym}} : \tensor{I} )^2
+
\left( 2 \mu + \kappa \right) \bar{\tensor{\varepsilon}}^{\text{sym}} : \bar{\tensor{\varepsilon}}^{\text{sym}} 
\right], \\
\label{eq:energy_parts_C}
& \psi_e^C ( \bar{\tensor{\varepsilon}}^{\text{skew}} ) =
\frac{1}{2} \kappa \bar{\tensor{\varepsilon}}^{\text{skew}} : \bar{\tensor{\varepsilon}}^{\text{skew}}, \\
\label{eq:energy_parts_R}
& \psi_e^R ( \bar{\tensor{\kappa}} ) = 
\frac{1}{2} 
\left[ 
\alpha ( \bar{\tensor{\kappa}} : \tensor{I} )^2
+
\beta \bar{\tensor{\kappa}} : \bar{\tensor{\kappa}}^{\text{T}}
+
\gamma \bar{\tensor{\kappa}} : \bar{\tensor{\kappa}}
\right].
\end{align} 
The force stress $\bar{\tensor{\sigma}}$ can be found by taking partial derivative of the energy density with respect to the micropolar strain. 
By using Eq.~\eqref{eq:strain_split1}, the force stress can also be partitioned into $\bar{\tensor{\sigma}}^B$ and $\bar{\tensor{\sigma}}^C$, which are the results of the pure non-polar deformation and the micro-continuum coupling effects, respectively:
\begin{equation}
\label{eq:force_stress}
\bar{\tensor{\sigma}} 
= 
\frac{\partial \psi_e}{\partial \bar{\tensor{\varepsilon}}} 
= 
\frac{\partial}{\partial \bar{\tensor{\varepsilon}}} ( \psi_e^B + \psi_e^C ) 
=
\underbrace{
\lambda ( \bar{\tensor{\varepsilon}}^{\text{sym}} : \tensor{I} ) \tensor{I}
+ \left( 2 \mu +\kappa \right) \bar{\tensor{\varepsilon}}^{\text{sym}}
}_{:= \bar{\tensor{\sigma}}^B }
+ 
\underbrace{
\kappa \bar{\tensor{\varepsilon}}^{\text{skew}}
}_{:= \bar{\tensor{\sigma}}^C}.
\end{equation}
Similarly, the couple stress $\bar{\tensor{m}}^R$ that is caused by the pure micro-rotation can be obtained as follows:
\begin{equation}
\label{eq:couple_stress}
\bar{\tensor{m}}^R 
= \frac{\partial \psi_e}{\partial \bar{\tensor{\kappa}}}
= \frac{\partial \psi_e^R}{\partial \bar{\tensor{\kappa}}} 
= \alpha ( \bar{\tensor{\kappa}} : \tensor{I} ) \tensor{I} + \beta \bar{\tensor{\kappa}}^\text{T} + \gamma \bar{\tensor{\kappa}}.
\end{equation}
Based on the split approach, notice that the partitioned energy densities in Eqs.~\eqref{eq:energy_parts_B}-\eqref{eq:energy_parts_R} can be recovered by:
\begin{equation}
\label{eq:energy_conjugates}
\psi_e^{B} = \frac{1}{2} \bar{\tensor{\sigma}}^B : \bar{\tensor{\varepsilon}}^{\text{sym}}
\: \: ; \: \:
\psi_e^{C} = \frac{1}{2} \bar{\tensor{\sigma}}^C : \bar{\tensor{\varepsilon}}^{\text{skew}}
\: \: ; \: \:
\psi_e^{R} = \frac{1}{2} \bar{\tensor{m}}^R : \bar{\tensor{\kappa}},
\end{equation}
where $(\bar{\tensor{\sigma}} = \bar{\tensor{\sigma}}^B + \bar{\tensor{\sigma}}^C, \bar{\tensor{\varepsilon}} = \bar{\tensor{\varepsilon}}^{\text{sym}} + \bar{\tensor{\varepsilon}}^{\text{skew}})$ and $(\bar{\tensor{m}}^R, \bar{\tensor{\kappa}})$ are the energy-conjugated pairs. 
This partition of energy then provides a mean to introduce different degradation mechanisms for different kinematic modes.

\section{Phase field model for damaged micropolar continua}
\label{phase_field}
This section presents a variational phase field framework to model cohesive fracture in micropolar materials. 
Our starting point is the energy split introduced in Section \ref{micropolar_elasticity}. 
We introduce distinct degradation for the Boltzmann, coupling and micro-rotational energy-conjugated pairs and derive, for the first time, the action functional for the variational phase field fracture framework for micropolar continua. 
The governing equations are then sought by seeking the stationary point, i.e., the Euler-Lagrange equation. 
To ensure that the size effect exhibited in the simulations are originated from the micropolar effect, we adopt the crack surface density functional originally proposed by \citet{wu2018length} and \citet{Geelen2019} to eliminate the sensitivity of the regularization length scale for the phase field fracture in Boltzmann continua. 
Our 1D analysis in Section \ref{1d_analysis} and numerical results in Section \ref{trapezoid} suggest that this same crack surface density functional may also eliminate the sensitivity of the regularization length scale parameter for the micropolar framework.

\subsection{Phase field approximation of cohesive fracture}
\label{phase_field_approximation}
This study adopts a phase field approach to represent cracks via an implicit function \citep{Bourdin2008, Miehe2010, Borden2012, clayton2015phase}. 
Let $\Gamma$ be the discontinuous surface within a micropolar elastic body $\mathcal{B}$. 
We approximate the fracture surface area $A_{\Gamma}$ as $A_{\Gamma_d}$, which is the volume integration of crack surface density $\Gamma_d \left( d, \grad{d} \right)$ over $\mathcal{B}$,
\begin{equation}
\label{eq:crack_surface}
A_{\Gamma} \approx A_{\Gamma_d} =\int_{\mathcal{B}} \Gamma_d \left( d, \grad{d} \right) \: dV,
\end{equation}
where $d$ is the phase field which varies from 0 in undamaged regions to 1 in completely damaged regions. 
In this study, we consider the following crack surface density functional, which is originally used to introduce elliptic regularization of the Mumford-Shah functional for image segmentation \citep{mumford1989optimal}, i.e., 
\begin{equation}
\label{eq:crack_density}
\Gamma_d \left( d, \grad{d} \right) = \frac{1}{c_0} \left[ \frac{1}{l_c} w ( d ) + l_c \left( \grad{d} \cdot \grad{d} \right) \right]
\: \: ; \: \:
c_0 = 4 \int_0^1 \sqrt{w (s)} \: ds,
\end{equation}
where $l_c$ is the regularization length scale that governs the size of the diffusive crack zone, $c_0>0$ is a normalization constant, and $w(d)$ is one of the function that controls the shape of the regularized profile of the phase field \citep{clayton2011phase,Mesgarnejad2015, bleyer2018phase}. 
$\Gamma-$convergence requires that the sharp cracks can be recovered by reducing the length scale parameter $l_{c}$ to zero such that \citep{clayton2015phase, wu2018length}, 
\begin{equation}
\label{eq:limit} A_{\Gamma} = \lim_{l_{c} \rightarrow 0} A_{\Gamma_d}.
\end{equation}
The local dissipation function $w(d)$ should be a monotonically increasing function of $d$, and we distinguish between the following two choices for this function:
\begin{equation}
\label{eq:local_energy}
w ( d ) 
=
\begin{dcases}
d^2 \\
d
\end{dcases}.
\end{equation}
The quadratic local dissipation $w(d) = d^2$ is the most widely used approach in simulating brittle fracture, and has become the standard in the phase field approximation \citep{Bourdin2008,Miehe2010,miehe2010thermodynamically,bourdin2011time,Borden2012, miehe2015phase2,miehe2015phase1,aldakheel2018phase,choo2018coupled,Bryant2018,chukwudozie2019variational}. 
The major disadvantage of the quadratic local dissipation is that the damage evolution initiates as soon as the load is applied so that there is no pure elastic response. 
However, the linear model $w(d) = d$, combined with suitable degradation functions, describes the cohesive fracture that possesses a threshold energy that is independent of the regularization length $l_c$ \citep{Lorentz2012, Mesgarnejad2015, lorentz2017nonlocal, Geelen2019}. 
Also, in this case, the material is characterized by an elastic phase until the stored energy density reaches the threshold value. 
Since this study aims to decouple the regularization length $l_c$ for large-scale simulations, we adopt the linear dissipation function to take advantage of the aforementioned characteristics. 
The expression for the crack surface energy density in Eq.~\eqref{eq:crack_density} then becomes:
\begin{equation}
\label{eq:crack_density_details}
\Gamma_d \left( d, \grad{d} \right) =
\frac{3}{8 l_c} d + \frac{3 l_c}{8} \left( \grad{d} \cdot \grad{d} \right).
\end{equation}

\subsection{Free energy functional}
\label{phase_field_free_energy_functional}
It should be noted that the crack propagation within a body $\mathcal{B}$ corresponds to the creation of new free surfaces $\Gamma$. 
This implies that the rate of change of the internal energy should be equal to the rate of change of the surface energy that contributes to the crack growth. 
By assuming that this concept can be applied to the micropolar elastic material as well, the total potential energy $\Psi$ can be defined as follows \citep{Bilgen2019, Geelen2019}:
\begin{equation}
\label{eq:potential_energy}
\Psi = \int_{\mathcal{B}} \psi_e ( \bar{\tensor{\varepsilon}}, \bar{\tensor{\kappa}} ) \: dV + \int_{\Gamma} \mathcal{G}_c \: d\Gamma,
\end{equation} 
where $\mathcal{G}_c$ is critical energy release rate that quantifies the resistance to cracking. 
Then, revisiting Eqs.~\eqref{eq:energy_split} and~\eqref{eq:crack_surface}, we approximate the functional by
\begin{equation}
\label{eq:potential_energy_approx_pre}
\Psi
\approx
\int_{\mathcal{B}} 
\psi
\: dV,
\end{equation}
with
\begin{equation}
\label{eq:potential_energy_approx}
\psi = 
\underbrace{
g_B (d) \psi_e^{B+} ( \bar{\tensor{\varepsilon}}^{\text{sym}} )
+
g_C (d) \psi_e^C ( \bar{\tensor{\varepsilon}}^{\text{skew}} )
+
g_R (d) \psi_e^R ( \bar{\tensor{\kappa}} )
+
\psi_e^{B-} ( \bar{\tensor{\varepsilon}}^{\text{sym}} )
}_{:= \psi_{\text{bulk}} ( \bar{\tensor{\varepsilon}}^{\text{sym}}, \bar{\tensor{\varepsilon}}^{\text{skew}}, \bar{\tensor{\kappa}}, d  ) }
+
\mathcal{G}_c \Gamma_d \left( d, \grad{d} \right),
\end{equation}
where $\psi_{\text{bulk}} ( \bar{\tensor{\varepsilon}}^{\text{sym}}, \bar{\tensor{\varepsilon}}^{\text{skew}}, \bar{\tensor{\kappa}}, d  )$ is the degrading elastic bulk energy, and $g_i (d)$ are the stiffness degradation functions for the corresponding fictitious undamaged energy density parts $\psi_e^i$ ($i = B, C, R$). 
Here, notice that we decompose $\psi_e^B ( \bar{\tensor{\varepsilon}}^{\text{sym}} )$ into a positive and negative parts, and degrade only the positive part in order to avoid crack propagation under compression, i.e.,
\begin{equation}
\label{eq:Boltzmann_decomposition1}
\psi_e^B = \psi_e^{B+} + \psi_e^{B-}.
\end{equation}
In this study, we adopt the spectral decomposition scheme of \citet{Miehe2010}, so that each part can be written as:
\begin{equation}
\label{eq:Boltzmann_decomposition2}
\psi_e^{B \pm} =
\frac{1}{2} 
\left[ \lambda \left\langle \bar{\tensor{\varepsilon}}^{\text{sym}} : \tensor{I} \right\rangle_{\pm}^2
+
\left( 2 \mu + \kappa \right) \bar{\tensor{\varepsilon}}_{\pm}^{\text{sym}} : \bar{\tensor{\varepsilon}}_{\pm}^{\text{sym}} 
\right]
\: \: ; \: \:
\bar{\tensor{\varepsilon}}_{\pm}^{\text{sym}} = \displaystyle\sum_{a = 1}^3 \left\langle \bar{\varepsilon}_a^{\text{sym}} \right\rangle_{\pm} \left( \vec{n}_a \otimes \vec{n}_a \right),
\end{equation}
where $\left\langle \bullet \right\rangle_{\pm} = \left( \bullet \pm \left| \bullet \right| \right)/2$ is the Macaulay bracket operator, $\bar{\varepsilon}_a^{\text{sym}}$ is the principal Boltzmann strains, and $\vec{n}_a$ are the corresponding principal directions.

In order to investigate the effects of each energy density part, one may assume that the partitioned strain energy densities can either be degraded $(i \in \mathfrak{D})$ or remain completely undamaged $(i \in \mathfrak{U})$, i.e.,
\begin{equation}
\label{eq:distinct_degradation}
g_i (d) =
\begin{dcases}
g (d) & \text{if } i \in \mathfrak{D} \\
1 & \text{if } i \in \mathfrak{U}
\end{dcases} 
\: \: ; \: \:
\mathfrak{D} \cup \mathfrak{U} = \left\lbrace B, C, R \right\rbrace 
\: \: ; \: \: 
\mathfrak{D} \cap \mathfrak{U} = \emptyset.
\end{equation}
Here, $g (d)$ is a monotonically decreasing function that satisfies the following conditions \citep{pham2013onset}:
\begin{equation}
\label{eq:degradation_restriction}
g ( 0 ) = 1 \: \: ; \: \: g ( 1 ) = 0 \: \: ; \: \: g' ( d ) \leq 0 \text{ for } d \in [0, 1],
\end{equation}
where the superposed prime denotes derivative with respect to $d$. 
Explicit form of this function is provided in Section \ref{crack_irreversibility}. 
Notice that this general approach can be tailored to many different situations. For example, one may only degrade the pure Boltzmann part of the strain energy, i.e., $\mathfrak{D} = \left\lbrace B \right\rbrace$, $\mathfrak{U} = \left\lbrace C, R \right\rbrace$:
\begin{equation}
\label{eq:damaging_B}
\psi_{\text{bulk}} ( \bar{\tensor{\varepsilon}}^{\text{sym}}, \bar{\tensor{\varepsilon}}^{\text{skew}}, \bar{\tensor{\kappa}}, d  ) = g (d) \psi_e^{B+} ( \bar{\tensor{\varepsilon}}^{\text{sym}} )
+ \psi_e^C ( \bar{\tensor{\varepsilon}}^{\text{skew}} )
+ \psi_e^R ( \bar{\tensor{\kappa}} )
+ \psi_e^{B-} ( \bar{\tensor{\varepsilon}}^{\text{sym}} ),
\end{equation}
or degrade the entire strain energy density, i.e., $\mathfrak{D} = \left\lbrace B, C, R \right\rbrace$, $\mathfrak{U} = \emptyset$:
\begin{equation}
\label{eq:damaging_all}
\psi_{\text{bulk}} ( \bar{\tensor{\varepsilon}}^{\text{sym}}, \bar{\tensor{\varepsilon}}^{\text{skew}}, \bar{\tensor{\kappa}}, d  ) = g (d) \left[ \psi_e^{B+} ( \bar{\tensor{\varepsilon}}^{\text{sym}} )
+ \psi_e^C ( \bar{\tensor{\varepsilon}}^{\text{skew}} )
+ \psi_e^R ( \bar{\tensor{\kappa}} ) \right]
+ \psi_e^{B-} ( \bar{\tensor{\varepsilon}}^{\text{sym}} ).
\end{equation}

\subsection{Derivation of Euler-Lagrange equations via variational principle}
\label{derivation_of_gov_eqs}
Let $\mathbb{V}$ denote an appropriate function space. 
Then, based on the fundamental lemma of calculus of variations, the necessary condition for the energy functional $\Psi : \mathbb{V} \to \mathbb{R}$ in Eq.~\eqref{eq:potential_energy_approx_pre} to have a local extremum at a point $\vec{\chi}_0 \in \mathbb{V}$ is that,
\begin{equation}
\label{eq:Euler_Lagrange_necessary}
\frac{\delta \psi}{\delta \vec{\chi}} (\vec{\chi}_0) 
= \vec{0},
\end{equation}
where $\psi$ is the energy density that is previously defined in Eq.~\eqref{eq:potential_energy_approx}, $\vec{\chi} := \left\lbrace \vec{u}, \vec{\theta}, d \right\rbrace$ indicates the field variables, and $\delta ( \bullet ) / \delta \vec{\chi}$ denotes the functional derivative with respect to $\vec{\chi}$. 
Notice that Eq.~\eqref{eq:Euler_Lagrange_necessary} is the so-called Euler-Lagrange equations, which yield the governing partial differential equations to be solved.

The linear momentum balance equation can be recovered by seeking the stationary point where the functional derivative of $\psi$ with respect to $\vec{u}$ vanishes. 
By assuming no body forces and by only considering the single derivative, we have, 
\begin{equation}
\label{eq:variational_BLM1}
\frac{\delta \psi}{\delta \vec{u}} 
= \frac{\partial \psi}{\partial \vec{u}}
- \diver{\frac{\partial \psi}{\partial \grad{\vec{u}}}} = \vec{0}.
\end{equation}
By revisiting Eq.~\eqref{eq:potential_energy_approx}, we get:
\begin{equation}
\label{eq:variational_BLM2}
\frac{\partial \psi}{\partial \vec{u}} = \vec{0},
\end{equation}
and by Eq.~\eqref{eq:force_stress},
\begin{equation}
\label{eq:variational_BLM3}
\diver{\frac{\partial \psi}{\partial \grad{\vec{u}}} } 
=
\diver{\left[ g_B(d) \frac{\partial \psi_e^B}{\partial \grad{\vec{u}}} + g_C(d) \frac{\partial \psi_e^C}{\partial \grad{\vec{u}}} \right]}
=
\diver{ \left[ g_B(d) \bar{\tensor{\sigma}}^B + g_C(d) \bar{\tensor{\sigma}}^C \right] },
\end{equation}
since the decomposition of the micropolar strain in Eq.~\eqref{eq:strain_split1} yields the following: 
\begin{align}
\label{eq:variational_BLM4}
&\frac{\partial \psi_e^B}{\partial \grad{\vec{u}}} 
= \frac{\partial \psi_e^B}{\partial \bar{\tensor{\varepsilon}}^{\text{sym}}} : \frac{\partial \bar{\tensor{\varepsilon}}^{\text{sym}}}{\partial \grad{\vec{u}}}
= \frac{\partial \psi_e^B}{\partial \bar{\tensor{\varepsilon}}^{\text{sym}}} : \frac{1}{2} \left( \mathbb{I} + \bar{\mathbb{I}} \right) = \frac{\partial \psi_e^B}{\partial \bar{\tensor{\varepsilon}}^{\text{sym}}} = \bar{\tensor{\sigma}}^B, \\
&\frac{\partial \psi_e^C}{\partial \grad{\vec{u}}} 
= \frac{\partial \psi_e^C}{\partial \bar{\tensor{\varepsilon}}^{\text{skew}}} : \frac{\partial \bar{\tensor{\varepsilon}}^{\text{skew}}}{\partial \grad{\vec{u}}} 
= \frac{\partial \psi_e^C}{\partial \bar{\tensor{\varepsilon}}^{\text{skew}}} : \frac{1}{2} \left( \bar{\mathbb{I}} - \mathbb{I} \right) 
= \frac{\partial \psi_e^C}{\partial \bar{\tensor{\varepsilon}}^{\text{skew}}} 
= \bar{\tensor{\sigma}}^C.
\end{align}
Similarly, assuming no body couples, the balance of angular momentum can be obtained by searching the local extremum where the functional derivative of $\psi$ with respect to the micro-rotation $\vec{\theta}$ vanishes, i.e.,
\begin{equation}
\label{eq:variational_BAM1}
\frac{\delta \psi}{\delta \vec{\theta}} 
= \frac{\partial \psi}{\partial \vec{\theta}}
- \diver{\frac{\partial \psi}{\partial \grad{\vec{\theta}}}} = \vec{0}.
\end{equation}
The partial derivative of $\psi$ with respect to $\vec{\theta}$ is:
\begin{equation}
\label{eq:variational_BAM2}
\frac{\partial \psi}{\partial \vec{\theta}}
= g_C(d) \frac{\partial \psi_e^C}{\partial \vec{\theta}} = g_C(d) \frac{\partial \psi_e^C}{\partial \bar{\tensor{\varepsilon}}^{\text{skew}}} : \frac{\partial \bar{\tensor{\varepsilon}}^{\text{skew}}}{\partial \vec{\theta}} 
= - \overset{3}{\tensor{E}} : \left[ g_C(d) \bar{\tensor{\sigma}}^C \right].
\end{equation}
By Eq.~\eqref{eq:couple_stress}, the partial derivative of $\psi$ with respect to $\grad{\vec{\theta}}$ becomes:
\begin{equation}
\label{eq:variational_BAM3}
\diver{\frac{\partial \psi}{\partial \grad{\vec{\theta}}}}
= \diver{ \left[ g_R (d) \frac{\partial \psi_e^R}{\partial \vec{\bar{\kappa}}} \right] } 
= \diver{ \left[ g_R (d) \bar{\tensor{m}}^R \right] }.
\end{equation}
The damage evolution equation (i.e., functional derivative of $\psi$ with respect to the phase field $d$) can also be recovered as follows:
\begin{equation}
\label{eq:variational_PF1}
\frac{\delta \psi}{\delta d}
= \frac{\partial \psi}{\partial d}
- \diver{\frac{\partial \psi}{\partial \grad{d}}} =0,
\end{equation}
where, by revisiting Eq.~\eqref{eq:crack_density_details},
\begin{equation}
\label{eq:variational_PF2}
\frac{\partial \psi}{\partial d} = g'(d) \left[ \displaystyle\sum_{i \in \mathfrak{D}} \psi_e^{i} \right] + \frac{3 \mathcal{G}_c}{8 l_c},
\end{equation}
and
\begin{equation}
\label{eq:variational_PF3}
\diver{\frac{\partial \psi}{\partial \grad{d}}} = \diver{\left( \frac{3 \mathcal{G}_c l_c}{4} \grad{d} \right)}.
\end{equation}
Finally, collecting the terms from Eqs.~\eqref{eq:variational_BLM1}-\eqref{eq:variational_PF3}, we obtain the following coupled system of partial differential equations to be solved: 
\begin{align}
\label{eq:strongform_BLM}
& \diver{ \left[ g_B(d) \bar{\tensor{\sigma}}^B + g_C(d) \bar{\tensor{\sigma}}^C \right] } = \vec{0} && \text{balance of linear momentum}, \\
\label{eq:strongform_BAM}
& \diver{ \left[ g_R(d) \bar{\tensor{m}}^R \right] } + \overset{3}{\tensor{E}} : \left[ g_C(d) \bar{\tensor{\sigma}}^C \right] = \vec{0} && \text{balance of angular momentum}, \\
\label{eq:strongform_PF}
& g' (d) \mathcal{F}
+ \frac{3}{8} \left( 1 - 2 l_c^2 \nabla^2 d \right) = 0 && \text{nondimensionalized damage evolution equation},
\end{align}
where $\nabla^2 ( \bullet ) = \diver{\grad{ ( \bullet ) }}$ is the Laplacian operator, and $\mathcal{F}$ is the degrading nondimensionalized strain energy density:
\begin{equation}
\label{eq:nondimensionalized_energy}
\mathcal{F} = \frac{\displaystyle\sum_{i \in \mathfrak{D}} \psi_e^{i}}{\mathcal{G}_c / l_c}.
\end{equation}

\subsection{Crack irreversibility and degradation function}
\label{crack_irreversibility}
As far as $\mathfrak{D} \neq \emptyset$, we prevent crack healing by following the treatment used in \citep{miehe2015minimization, choo2018coupled, Bryant2018} which ensures the irreversibility constraint by enforcing the driving force to be non-negative. 
Although the stored energy density is split into three different parts, we simply introduce one distinct history function or driving force $\mathcal{H}$ which is the pseudo-temporal maximum of the degrading nondimensionalized energy density. 
Inserting our definition into Eq.~\eqref{eq:strongform_PF} gives:
\begin{equation}
\label{eq:crack_irreversibility}
g' (d) \mathcal{H}
+ \frac{3}{8} \left( 1 - 2 l_c^2 \nabla^2 d \right) = 0.
\end{equation}
Revisiting Section ~\ref{phase_field_approximation}, the term cohesive denotes that the model should possess a threshold for the loading, where damage does not develop below this value. 
Therefore, we particularly restrict the crack growth to initiate above a threshold energy density $\psi_{\text{crit}}$ by using the following history function, in order to approximate the cohesive response:
\begin{equation}
\label{eq:crack_driving_force}
\mathcal{H} = \max_{\tau \in [0, t]} \left\lbrace \mathcal{F}_{\text{crit}} + \mathcal{F}_{\text{crit}} \left\langle \frac{\mathcal{F}}{\mathcal{F}_{\text{crit}}} - 1 \right\rangle_{+}  \right\rbrace 
\: \: ; \: \:
\mathcal{F}_{\text{crit}} = \frac{\psi_{\text{crit}}}{\mathcal{G}_c / l_c},
\end{equation}
where $\mathcal{F}_{\text{crit}}$ is the nondimensionalized threshold energy. 
Note that Eq.~\eqref{eq:crack_irreversibility} is the field equation that is actually solved for the phase field in this study.

We complete our formulation by specifying the degradation function $g (d)$. 
This study adopts a quasi-quadratic degradation function \citep{lorentz2011convergence,Lorentz2012,Geelen2019}, which is a rational function of the phase field $d$. 
The quasi-quadratic degradation function has an associated upper bound on the regularization length $l_c$, and is defined as:
\begin{equation}
\label{eq:degradation_function}
g (d) = \frac{\left( 1 - d \right)^2}{\left( 1 - d \right)^2 + m d \left(1 + p d \right) }
\: \: ; \: \:
l_c \leq \frac{3 \mathcal{G}_c}{8 \left( p + 2 \right) \psi_{\text{crit}}},
\end{equation}
where $m \geq 1$ is constant, and $p \geq 1$ is a shape parameter that controls the peak stress and the fracture responses.

Recall that this study restricts the damage evolution to initiate above a threshold energy. 
In other words, below the threshold (i.e., $\mathcal{F} / \mathcal{F}_{\text{crit}} \leq 1$), the driving force and the phase field should satisfy $\mathcal{H} = \mathcal{F}_{\text{crit}}$ and $d = 0$, respectively. 
In this case, the damage evolution equation [Eq.~\eqref{eq:crack_irreversibility}] becomes:
\begin{equation}
\label{eq:damage_evolution_below_threshold}
g ' (0) \mathcal{F}_{\text{crit}} + \frac{3}{8} = 0.
\end{equation}
Since the degradation function [Eq.~\eqref{eq:degradation_function}] yields $g'(0) = -m$, we require 
\begin{equation}
\label{eq:damage_evolution_m_value}
m = \frac{3}{8 \mathcal{F}_{\text{crit}}},
\end{equation}
in order to trivially satisfy Eq.~\eqref{eq:damage_evolution_below_threshold}. 
Again, the most common choice for the degradation function would be a simple quadratic function, i.e., $g(d) = (1 - d)^2$. 
A phase field model that adopts the quadratic degradation function requires a particular critical energy $\psi_{\text{crit}}$ that depends on $l_c$ to satisfy Eq.~\eqref{eq:damage_evolution_below_threshold}. 
However, by using the quasi-quadratic degradation function in Eq.~\eqref{eq:degradation_function} with Eq.~\eqref{eq:damage_evolution_m_value}, it is noted that the threshold energy density $\psi_{\text{crit}}$ is no longer dependent on $l_c$, since the degradation function itself is designed to automatically satisfy Eq.~\eqref{eq:damage_evolution_below_threshold}. 
It reveals that the elastic response (i.e., if stored energy is below the threshold) is regularization length independent. 
The regularization length insensitivity of the model, for the case where the stored energy exceeds the threshold, will be discussed in Section \ref{1d_analysis}.

\subsection{One-dimensional analysis on phase field regularization length sensitivity}
\label{1d_analysis}
In order to gain insights on the regularization length insensitive response, we consider a similiar 1D boundary value problem previously used in \citet{wu2018length} and \citet{Geelen2019} for length scale analysis. 
Our major departure is that the material is now an analog to the micropolar material where an length scale dependent state variable (which replaces the micro-rotation due to the low-dimensional kinematics) is introduced to replicate the size effect of the elasticity response. 
Consider a one-dimensional bar $x \in [-L, L]$ subjected to a tensile loading on both ends. 
We assume that the length $2L$ is sufficiently long enough so that the any possible boundary effects can be neglected. 
We define the strain measures as $\bar{\varepsilon} = \mathrm{d} u/ \mathrm{d} x$, and $\bar{\kappa} = l_e (\mathrm{d} \theta/ \mathrm{d} x)$, where $l_e$ is the length scale. 
Our goal here is to check whether the size-dependent responses in the damaged zone is sensitive to the regularization length scale for the phase field $l_{c}$. 
Note that this formulation does yield size-dependent responses in both elastic and damage zones, but the kinematics in 1D does not permit rotation. 
Hence, $\theta$ and $\bar{\kappa}$ no longer indicate micro-rotation and micro-curvature respectively. 
Thus, analogous to Eq.~\eqref{eq:energy_split}, we introduce a 1D size-dependent model of which the strain energy density takes a quadratic form as:
\begin{equation}
\label{eq:1d_strainenergy}
\psi_e 
= \frac{1}{2} C_B \left( \frac{\mathrm{d} u}{\mathrm{d} x} \right)^2 
+ \frac{1}{2} C_C \left( \frac{\mathrm{d} u}{\mathrm{d} x} - l_e \frac{\mathrm{d} \theta}{\mathrm{d} x} \right)^2 
+ \frac{1}{2} C_R l_e^2 \left( \frac{\mathrm{d} \theta}{\mathrm{d} x} \right)^2,
\end{equation}
where $C_B >0$, $C_C >0$ and $C_R >0$ are the material parameters. 
Notice that (1) this stored energy functional does not admit non-trivial zero-energy mode provided that $C_{B} + C_{C} > 0$ and $C_{R}> 0$ and 
(2) if the length scale $l_e$ vanishes, Eq.~\eqref{eq:1d_strainenergy} reduces to an energy functional for the classical Boltzmann continuum. 
In this setting, the stress measures can be obtained as,
\begin{align}
\label{eq:1d_stresses_1}
&\bar{\sigma} = \frac{\partial \psi_e}{\partial \bar{\varepsilon}} = C_B \bar{\varepsilon} + C_C \left( \bar{\varepsilon} - \bar{\kappa} \right), \\
\label{eq:1d_stresses_2}
&\bar{m} = \frac{\partial \psi_e}{\partial \bar{\kappa}} = C_R \bar{\kappa} - C_C \left( \bar{\varepsilon} - \bar{\kappa} \right),
\end{align}
where both $\bar{\sigma}$ and $\bar{m}$ are $l_e$-dependent. 
Assuming all energy density parts can be degraded, the Lagrangian for the damaged state where $\psi_e > \psi_{\text{crit}}$ then becomes:
\begin{equation}
\label{eq:1d_totalenergy}
\psi = g(d) \psi_e
+ \frac{3 \mathcal{G}_c}{8 l_c} \left[ d + l_c^2 \left( \frac{\mathrm{d} d}{\mathrm{d} x} \right)^2 \right],
\end{equation}
where $g(d)$ is previously defined in Eq.~\eqref{eq:degradation_function}. 
The first variation of Eq.~\eqref{eq:1d_totalenergy} yields the following set of Euler-Lagrange equations:
\begin{align}
\label{eq:1d_BLM}
&\frac{\mathrm{d}}{\mathrm{d} x} \left[ g(d) \bar{\sigma} \right] = 0, \\
\label{eq:1d_BAM}
&\frac{\mathrm{d}}{\mathrm{d} x} \left[ g(d) \bar{m} \right]  = 0,  \\
\label{eq:1d_PF}
&g'(d) \frac{\psi_e}{\mathcal{G}_c / l_c}
+
\frac{3}{8} \left( 1 - 2 l_c^2 \frac{\mathrm{d}^2 d}{\mathrm{d} x^2} \right) = 0,
\end{align}
where Eqs.~\eqref{eq:1d_BLM} and~\eqref{eq:1d_BAM} are the balance equations, and Eq.~\eqref{eq:1d_PF} is the nondimensionalized damage evolution equation. 
Following \citet{Geelen2019}, we apply a specific amount of $\bar{\sigma}$ at both ends while the boundaries remain $\bar{m}$-free, such that Eq.~\eqref{eq:1d_BLM} and Eq.~\eqref{eq:1d_BAM} yields:
\begin{equation}
\label{eq:1d_solvingBLMandBAM}
g(d) \bar{\sigma} = \bar{\sigma}_0
\: \: ; \: \:
g(d) \bar{m} = 0,
\end{equation}
where $\bar{\sigma}_0$ is the responding stress on the boundary ($x = \pm L$). 
By the constitutive relationships in Eqs.~\eqref{eq:1d_stresses_1}-~\eqref{eq:1d_stresses_2}, we can now express the strain measures $\bar{\varepsilon}$ and $\bar{\kappa}$ as,
\begin{equation}
\label{eq:strainmeasures}
\bar{\varepsilon} = \frac{C_C - C_R}{C_B (C_C - C_R) - C_C C_R} \frac{\bar{\sigma}_0}{g(d)}
\: \: ; \: \:
\bar{\kappa} = \frac{C_C}{C_B (C_C - C_R) - C_C C_R} \frac{\bar{\sigma}_0}{g(d)}.
\end{equation}
Thus, the energy density functional in Eq.~\eqref{eq:1d_strainenergy} can be rewritten as:
\begin{equation}
\label{eq:1d_strainenergy2}
\psi_e = \frac{\bar{\sigma}_0^2}{C^* g(d)^2}
\: \: ; \: \:
C^* = \frac{2 \left[ C_B (C_C - C_R) - C_C C_R \right]^2}{C_B (C_C - C_R)^2 + C_C C_R (C_C + C_R)}.
\end{equation}
Substituting Eq.~\eqref{eq:1d_solvingBLMandBAM} and Eq.~\eqref{eq:1d_strainenergy2} into Eq.~\eqref{eq:1d_PF}, we get, 
\begin{equation}
\label{eq:1d_PF2}
g'(d)\left[ \frac{1}{\mathcal{G}_c / l_c} \frac{\bar{\sigma}_0^2}{C^* g(d)^2} \right] 
+
\frac{3}{8} \left( 1 - 2 l_c^2 \frac{\mathrm{d}^2 d}{\mathrm{d} x^2} \right) = 0.
\end{equation}
By multiplying Eq.~\eqref{eq:1d_PF2} with $\mathrm{d} d / \mathrm{d} x$, we can obtain the following differential equation: 
\begin{equation}
\label{eq:1d_PF3}
\frac{\mathrm{d}}{\mathrm{d} x}
\left\lbrace 
\frac{3}{8} \left[ d - l_c^2 \left( \frac{\mathrm{d} d}{\mathrm{d} x} \right)^2 \right] - \frac{1}{\mathcal{G}_c / l_c} \frac{\bar{\sigma}_0^2}{2 C^* g(d)}
\right\rbrace = 0.
\end{equation}
Following \citet{wu2018length} and \citet{Geelen2019}, we focus on the damaged zone $[-l_z, l_z]$ with $l_z \ll L$, where the outer edge of the zone is related to the parameter $d^*$, the maximum value of the damage across the bar (i.e., $d^* = 1$ if the bar is fully broken). 
Then, by symmetry, at $x=0$ we have:
\begin{equation}
\label{eq:1d_BC1}
d(d^*, 0) = d^*
\: \: ; \: \:
\frac{\mathrm{d} d}{\mathrm{d} x} (d^*, 0) = 0, 
\end{equation}
while the boundary conditions at the outer edge $x = l_z$ are given by,
\begin{equation}
\label{eq:1d_BC2}
d(d^*, l_z) = 0
\: \: ; \: \:
\frac{\mathrm{d} d}{\mathrm{d} x} (d^*, l_z) = 0.
\end{equation}
By introducing the parameter $d^*$, notice that our goal is to find the expression for the responding stress $\bar{\sigma}_0$ as a function of the maximum damage $d^*$, or vice versa. 
Integration of Eq.~\eqref{eq:1d_PF3}, using boundary conditions in Eq.~\eqref{eq:1d_BC2} admits the following:
\begin{equation}
\label{eq:1d_PF4}
\frac{3}{8} \left[ d - l_c^2 \left( \frac{\mathrm{d} d}{\mathrm{d} x} \right)^2 \right]
=
\frac{1}{\mathcal{G}_c / l_c} \frac{\bar{\sigma}_0^2}{2 C^*} \left( g(d)^{-1} - 1\right).
\end{equation}
Applying the symmetry conditions [Eq.~\eqref{eq:1d_BC1}], Eq.~\eqref{eq:1d_PF4} becomes:
\begin{equation}
\label{eq:1d_PF5}
\frac{\bar{\sigma}_0^2}{2 C^*}
= \frac{3 \mathcal{G}_c}{8 l_c} \frac{d^*}{g(d^*)^{-1} - 1}.
\end{equation}
Substituting the expression of the degradation function in Eq.~\eqref{eq:degradation_function} and the expression for the degradation parameter $m$ in Eq.~\eqref{eq:damage_evolution_m_value}, we finally get:
\begin{equation}
\label{eq:1d_PF6}
\bar{\sigma}_0 
=
\sqrt{
2 C^* \psi_{\text{crit}} \frac{(1-d^*)^2}{1 + p d^*}
}.
\end{equation}
Observe that the resultant stress $\bar{\sigma}_0$ can be expressed in terms of $d^*$ but independent of the phase field regularization length $l_c$. 
Eq.~\eqref{eq:1d_PF6} highlights that the proposed model is capable of replicating the global response insensitive to regularization length $l_c$ for the phase field, while preserving the size effect introduced by the micropolar elasticity. 
This result is important, as this insensitivity to $l_c$ enables us to simulate cohesive fracture in large spatial domain composed of micropolar materials. 
Extending this analysis for 2D and 3D cases is out of the scope of this study but will be considered in the future. 

\rmk{Previous works on phase field and gradient damage models for cohesive fracture in Cauchy continuum, such as \citet{cazes2010cohesive, Lorentz2012, wu2018geometrically}, have established a connection between the cohesive zone models and the phase field and gradient damage models that represent cracks via implicit function. 
In principle, it is possible that similar connection can be established between the micropolar phase field model presented in this paper and the established micropolar cohesive zone models such as \citet{larsson2007homogenization, zhang2007interface, hirschberger2008computational}. 
Such an endeavor is obviously out of the scope of this study due to the extensive length, but will be considered in future studies.}

\section{Finite element implementation}
\label{FE_implementation}
In this section, we describe the finite element discretization, followed by the solution strategy to solve the system of nonlinear equation incrementally. 
Starting from the strong form, we follow the standard procedure to recover the variational form while employing the Taylor-Hood finite element space for the displacement and micro-rotation fields, and standard linear interpolation for the phase field. 
This finite element space is chosen to match the design of the operator-split algorithm. 
The displacement and micro-rotation are updated in a monolithic manner, where we use Taylor-Hood element such that the displacement field is interpolated by quadratic polynomials and the micro-rotation is interpolated by linear polynomials. 
Meanwhile, the phase field is also interpolated by linear function to ensure the efficiency of the staggered solver that updates the phase field while holding the displacement and micro-rotation fixed. 
The operator-split solution scheme (i.e., staggered scheme) that successively updates the field variables is described in Section \ref{operator_split}.

\subsection{Galerkin form}
\label{Galerkin_form}
We derive the weak form and introduce the finite dimensional space to introduce the numerical scheme for the boundary value problems described in Eqs.~\eqref{eq:strongform_BLM}-\eqref{eq:strongform_BAM} and Eq.~\eqref{eq:crack_irreversibility}. 
We consider a micropolar elastic domain $\mathcal{B}$ with boundary $\partial \mathcal{B}$ composed of Dirichlet boundaries (displacement $\partial \mathcal{B}_u$ and micro-rotation $\partial \mathcal{B}_{\theta}$) and Neumann boundaries (traction $\partial \mathcal{B}_{t_{\sigma}}$ and moment $\partial \mathcal{B}_{t_{m}}$) satisfying,
\begin{equation}
\label{eq:boundary_decomposition}
\partial \mathcal{B} 
= 
\overline{\partial \mathcal{B}_u \cup \partial \mathcal{B}_{t_{\sigma}}}
=
\overline{\partial \mathcal{B}_{\theta} \cup \partial \mathcal{B}_{t_{m}}}
\: \: ; \: \:
\emptyset
= 
\partial \mathcal{B}_u \cap \partial \mathcal{B}_{t_{\sigma}}
=
\partial \mathcal{B}_{\theta} \cap \partial \mathcal{B}_{t_{m}}.
\end{equation}
The prescribed boundary conditions can be specified as:
\begin{align}
\label{eq:BCs}
\begin{dcases}
\vec{u} = \hat{\vec{u}} & \text{on } \partial \mathcal{B}_u, \\
\vec{\theta} = \hat{\vec{\theta}} & \text{on } \partial \mathcal{B}_{\theta}, \\
\left[ g_B ( d ) \bar{\tensor{\sigma}}^B + g_C ( d ) \bar{\tensor{\sigma}}^C \right]^{\text{T}} \cdot \vec{n} = \hat{\vec{t}}_{\sigma} & \text{on } \partial \mathcal{B}_{t_{\sigma}}, \\
\left[ g_R ( d ) \bar{\tensor{m}}^R \right]^{\text{T}} \cdot \vec{n} = \hat{\vec{t}}_{m} & \text{on } \partial \mathcal{B}_{t_{m}}, \\
\grad{d} \cdot \vec{n} = 0 & \text{on } \partial \mathcal{B},
\end{dcases}
\end{align}
where $\vec{n}$ is the outward-oriented unit normal on the boundary surface $\partial \mathcal{B}$; $\hat{\vec{u}}, \hat{\vec{\theta}}, \hat{\vec{t}}_{\sigma}, \hat{\vec{t}}_{m}$ are the prescribed displacement, micro-rotation, traction and moment, respectively; and the degradation functions $g_i (d)$ ($i = B, C, R$) are previously defined in Eq.~\eqref{eq:distinct_degradation}. 
For the model closure, the initial conditions are imposed as,
\begin{equation}
\label{eq:ICs}
 \vec{u} = \vec{u}_0 \: \: ; \: \: \vec{\theta} = \vec{\theta}_0,
\end{equation}
at $t = 0$.

We define the trial spaces $V_u$, $V_{\theta}$, and $V_d$ for the solution variables:
\begin{align}
\label{eq:trial_space_u}
&V_u = \left\lbrace \vec{u} : \mathcal{B} \to \mathbb{R}^3 \: | \: \vec{u} \in [ H^1 ( \mathcal{B} ) ]^3 , \: \left. \vec{u} \right|_{\partial \mathcal{B}_u} = \hat{\vec{u}} \right\rbrace, \\
\label{eq:trial_space_theta}
&V_{\theta} = \left\lbrace \vec{\theta} : \mathcal{B} \to \mathbb{R}^3 \: | \: \vec{\theta} \in [ H^1 ( \mathcal{B} ) ]^3 , \: \left. \vec{\theta} \right|_{\partial \mathcal{B}_{\theta}} = \hat{\vec{\theta}} \right\rbrace, \\
\label{eq:trial_space_d}
&V_d = \left\lbrace d : \mathcal{B} \to \mathbb{R} \: | \: d \in H^1 ( \mathcal{B} ) \right\rbrace,
\end{align}
where $H^1$ denotes the Sobolev space of order 1. 
Notice that this study adopts Taylor-Hood finite element (i.e., quadratic interpolation for displacement and linear for micro-rotation) following \citet{verhoosel2013phase}, which showed that the cohesive fracture model exhibits stress oscillation when equal order polynomials are used for the solution field, while the discretization with high order interpolation function for the displacement and first order functions for the auxiliary field and the phase field seems to eliminate this oscillation. 
Similarly, the corresponding admissible spaces for Eqs.~\eqref{eq:trial_space_u}-\eqref{eq:trial_space_d} with homogeneous boundary conditions are defined as,
\begin{align}
\label{eq:weighting_space_u}
&V_{\eta} = \left\lbrace \vec{\eta} : \mathcal{B} \to \mathbb{R}^3 \: | \: \vec{\eta} \in [ H^1 ( \mathcal{B} ) ]^3 , \: \left. \vec{\eta} \right|_{\partial \mathcal{B}_u} = \vec{0} \right\rbrace, \\
\label{eq:weighting_space_theta}
&V_{\xi} = \left\lbrace \vec{\xi} : \mathcal{B} \to \mathbb{R}^3 \: | \: \vec{\xi} \in [ H^1 ( \mathcal{B} ) ]^3 , \: \left. \vec{\xi} \right|_{\partial \mathcal{B}_{\theta}} = \vec{0} \right\rbrace, \\
\label{eq:weighting_space_d}
&V_{\zeta} = \left\lbrace \zeta : \mathcal{B} \to \mathbb{R} \: | \: \zeta \in H^1 ( \mathcal{B} ) \right\rbrace.
\end{align}
Applying the standard weighted residual procedure, the weak statements for Eqs.~\eqref{eq:strongform_BLM}-\eqref{eq:strongform_BAM} and Eq.~\eqref{eq:crack_irreversibility} is to: find $\left\lbrace \vec{u}, \vec{\theta}, d \right\rbrace \in V_u \times V_{\theta} \times V_d$ such that for all $\left\lbrace \vec{\eta},\vec{\xi},\zeta \right\rbrace \in V_{\eta} \times V_{\xi} \times V_{\zeta}$,
\begin{equation}
\label{eq:weak_form}
G_{u} ( \vec{u}, \vec{\theta}, d, \vec{\eta} ) = G_{\theta} ( \vec{u}, \vec{\theta}, d, \vec{\xi} ) = G_d ( \vec{u}, \vec{\theta}, d, \zeta )= 0.
\end{equation}
Here, $G_u \to \mathbb{R}$ is the weak statement of the balance of linear momentum:
\begin{equation}
\label{eq:weakform_BLM}
G_u = 
\int_{\mathcal{B}} \grad{\vec{\eta}} : \left[ g_B ( d ) \bar{\tensor{\sigma}}^B + g_C ( d ) \bar{\tensor{\sigma}}^C \right] \: dV
- 
\int_{\partial \mathcal{B}_{t_{\sigma}}} \vec{\eta} \cdot \hat{\vec{t}}_{\sigma} \: dA
= 0,
\end{equation}
$G_{\theta} \to \mathbb{R}$ is the weak statement of the balance of angular momentum:
\begin{equation}
\label{eq:weakform_BAM}
G_{\theta}
=
\int_{\mathcal{B}} \grad{\vec{\xi}} : g_R ( d ) \bar{\tensor{m}}^R \: dV
-
\int_{\mathcal{B}} \vec{\xi} \cdot \overset{3}{\tensor{E}} : \left[ g_C ( d ) \bar{\tensor{\sigma}}^C \right] \: dV
-
\int_{\partial \mathcal{B}_{t_{m}}} \vec{\xi} \cdot \hat{\vec{t}}_{m} \: dA
= 0,
\end{equation}
and $G_d \to \mathbb{R}$ is the weak statement of the damage evolution equation:
\begin{equation}
\label{eq:weakform_PF}
G_d
=
\int_{\mathcal{B}} \zeta \cdot g' (d) \mathcal{H} \: dV
+
\frac{3}{8} \left[
\int_{\mathcal{B}} \zeta + 2 l_c^2 \left( \grad{\zeta} \cdot \grad{d} \right) \: dV \right] = 0,
\end{equation}
where $\mathcal{H}$ and $g(d)$ are previously defined in Eqs.~\eqref{eq:crack_driving_force} and \eqref{eq:degradation_function}, respectively.

\subsection{Operator-split solution scheme}
\label{operator_split}
As previous studies on the phase field model showed that the operator splitting (i.e., staggered scheme) may potentially be more robust compared to the monolithic approach \citep{Miehe2010, heister2015primal, teichtmeister2017phase}, this study adopts the solution procedure based on the operator-split scheme to successively update three field variables $\left\lbrace \vec{u}, \vec{\theta}, d \right\rbrace$. 
In this operator-split setting, the damage field is updated first while the displacement and micro-rotation fields are held fixed. 
A new damage field $d_{\text{n+1}}$ is obtained iteratively once the algorithm converges within a predefined tolerance. 
Then, the linear solver holds the damage field fixed and advances the displacement and the micro-rotation fields $\left\lbrace \vec{u}_{\text{n+1}}, \vec{\theta}_{\text{n+1}} \right\rbrace$. 
The schematic of the solution strategy can be summarized as follows:
\begin{equation}
\label{eq:solution_strategies}
\begin{bmatrix}
\vec{u}_{\text{n}} \\
\vec{\theta}_{\text{n}} \\
d_{\text{n}}
\end{bmatrix} \underbrace{
\xrightarrow{\mathcal{R}(d) = 0}
\begin{bmatrix}
\vec{u}_{\text{n}} \\
\vec{\theta}_{\text{n}} \\\
d_{\text{n+1}}
\end{bmatrix}}_{\text{Iterative solver}}
\overbrace{
\xrightarrow{\mathcal{R}(\vec{u},\vec{\theta}) = \vec{0}}
\begin{bmatrix}
\vec{u}_{\text{n+1}} \\
\vec{\theta}_{\text{n+1}} \\
d_{\text{n+1}}
\end{bmatrix}}^{\text{Linear solver}},
\end{equation}
where $\mathcal{R}(\vec{u},\vec{\theta})$ and $\mathcal{R}(d)$ are the residuals that are consistent with Eqs.~\eqref{eq:weakform_BLM}-\eqref{eq:weakform_PF}:
\begin{align}
\label{eq:residuals}
\mathcal{R}(\vec{u},\vec{\theta})
&:
\begin{dcases}
\int_{\mathcal{B}} \grad{\vec{\eta}} : \left[ g_B ( d_{\text{n+1}} ) \bar{\tensor{\sigma}}_{\text{n+1}}^B + g_C ( d_{\text{n+1}} ) \bar{\tensor{\sigma}}_{\text{n+1}}^C \right] \: dV
- 
\int_{\partial \mathcal{B}_{t_{\sigma}}} \vec{\eta} \cdot \left. \hat{\vec{t}}_{\sigma} \right|_{\text{n+1}} \: dA, \\
\int_{\mathcal{B}} \grad{\vec{\xi}} : g_R ( d_{\text{n+1}} ) \bar{\tensor{m}}_{\text{n+1}}^R \: dV
-
\int_{\mathcal{B}} \vec{\xi} \cdot \overset{3}{\tensor{E}} : \left[ g_C ( d_{\text{n+1}} ) \bar{\tensor{\sigma}}_{\text{n+1}}^C \right] \: dV
-
\int_{\partial \mathcal{B}_{t_{m}}} \vec{\xi} \cdot \left. \hat{\vec{t}}_{m} \right|_{\text{n+1}} \: dA
\end{dcases} \\[1.25ex]
\mathcal{R}(d)
&:
\begin{dcases}
\int_{\mathcal{B}} \zeta \cdot g' (d_{\text{n+1}}) \mathcal{H}_{\text{n}} \: dV
+
\frac{3}{8} \left[
\int_{\mathcal{B}} \zeta + 2 l_c^2 \left( \grad{\zeta} \cdot \grad{d}_{\text{n+1}} \right) \: dV \right] 
\end{dcases}.
\end{align}
It should be noticed that one may choose other strategies to solve the same system of equations, however, the exploration of different schemes are out of the scope of this study.

The implementation of the numerical models including the finite element discretization and the operator-split solution scheme rely on the finite element package \verb|FEniCS| \citep{logg2010dolfin,logg2012automated,logg2012dolfin,alnaes2015fenics} with \verb|PETSc| scientific computation toolkit \citep{abhyankar2018petsc}. 
The scripts developed for this study are open-sourced (available at \url{https://github.com/hyoungsuksuh/micropolar_phasefield}), in order to aid third-party verification and validation \citep{suh2019open}.

\section{Numerical examples}
\label{examples}
This section presents numerical examples to showcase the applicability of the proposed phase field model for damaged micropolar elastic material. 
For simplicity, we limit our attention to two-dimensional simulations in this section. 
Based on the 2D setting, the kinematic state of the micropolar elastic body can be described by two in-plane displacements $\vec{u} = [u_1, u_2]^{\text{T}}$ and one out-of-plane micro-rotation angle $\theta_3$. 
Since the material elasticity now only depends on bending characteristic length, we now require only four engineering material parameters (e.g., $E$, $\nu$, $N$, and $l_b$).

The first example serves as a verification test that highlights the regularization length insensitive response of the phase field model with quasi-quadratic degradation function in Eq.~\eqref{eq:degradation_function}. 
We then investigate the effect of scale-dependent elasticity on the crack patterns, by simulating asymmetric notched three-point bending tests with different coupling numbers $N$ and single edge notched tests with different characteristic lengths $l_b$, respectively. 
Finally, we exhibit the applicability of the proposed energy split scheme by considering different degradation functions on the partitioned energy densities. 
All the numerical simulations rely on meshes that are sufficiently refined to properly capture the damage field around crack surfaces. 
Unless specified, we especially adopt the element size of $h^e \approx l_c/10$ around the potential crack propagation trajectory.

\subsection{Verification exercise: the trapezoid problem}
\label{trapezoid}
\begin{figure}[h]
\centering
\includegraphics[width=0.7\textwidth]{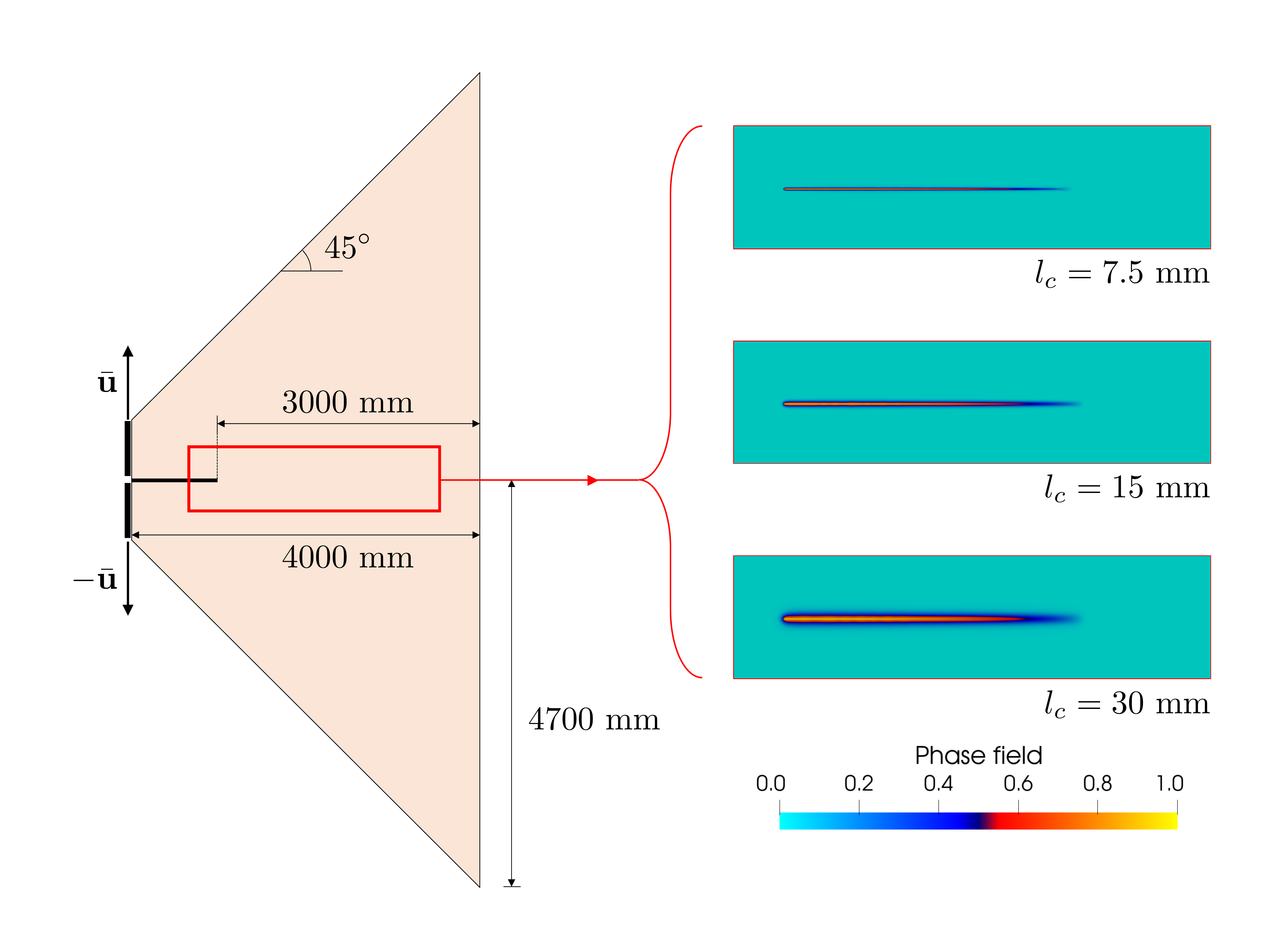}
\caption{Schematic of geometry and boundary conditions for a trapezoidal domain and the observed crack patterns at $\bar{u} = 0.325$ mm with different regularization lengths.}
\label{fig:trapezoid_domain}
\end{figure}

We first examine a problem proposed by \citet{Lorentz2012} that has a trapezoidal-shaped symmetrical domain with an initial notch. 
Since we prescribe the displacement $\vec{\bar{u}}$ in order to consider pure Mode I loading, as illustrated in Fig.~\ref{fig:trapezoid_domain}, this specific geometry helps us to avoid crack kinking and to facilitate straight crack propagation. 
The aforementioned characteristics of the trapezoid problem makes it suitable for verifying the regularization length insensitive response of the cohesive phase field model with quasi-quadratic degradation. 
This example thus performs a parametric study for three different regularization lengths ($l_c =$ 7.5, 15, and 30 mm, as depicted in Fig.~\ref{fig:trapezoid_domain}), with two different shape parameters $p =$ 2.5 and 10. 

\begin{figure}[h]
\centering
\subfigure[Shape parameter $p = 2.5$.]{\label{fig:trapezoid_p025}
\includegraphics[height=0.375\textwidth]{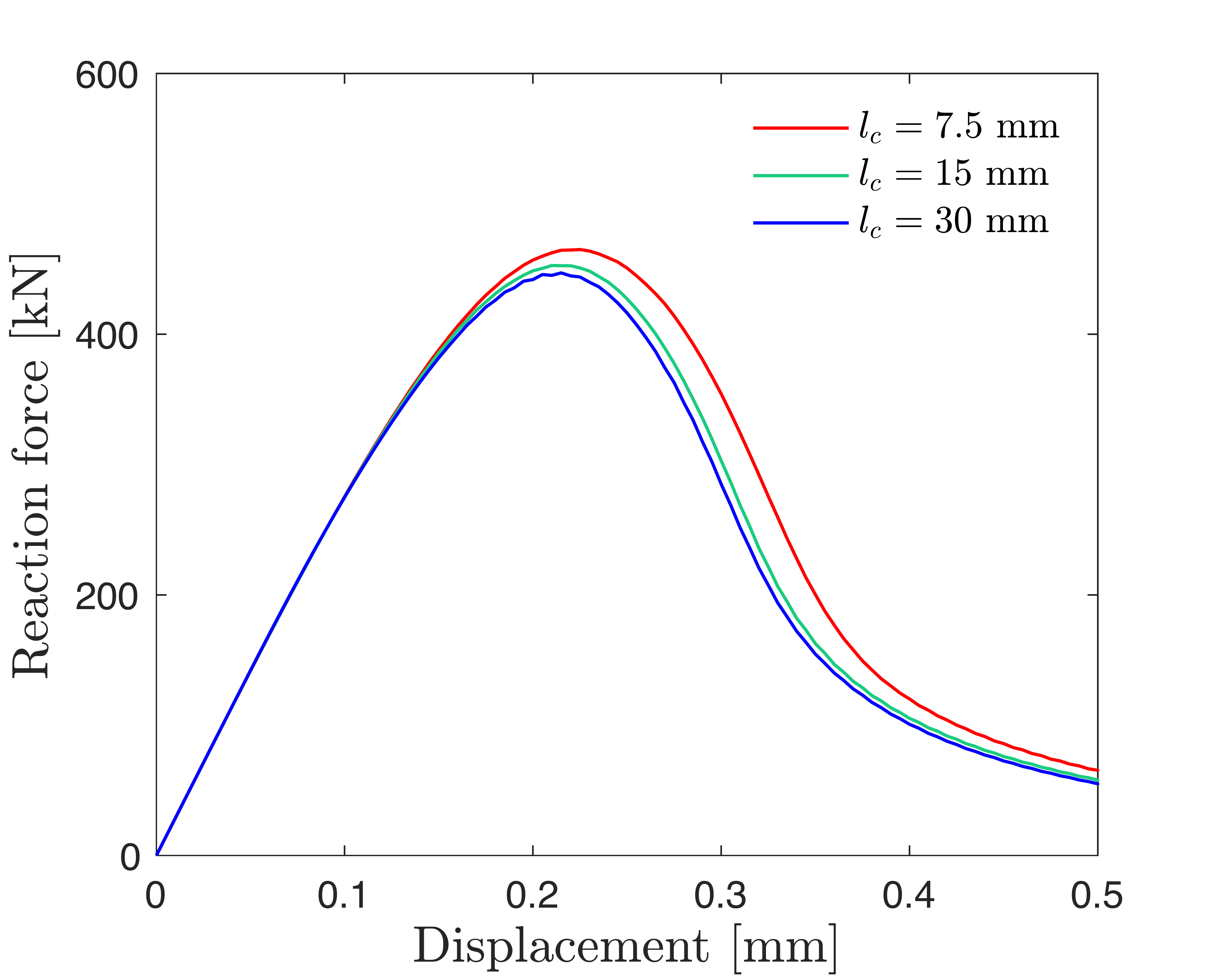}}
\hspace{0.01\textwidth}
\subfigure[Shape parameter $p = 10$.]{\label{fig:trapezoid_p100}
\includegraphics[height=0.375\textwidth]{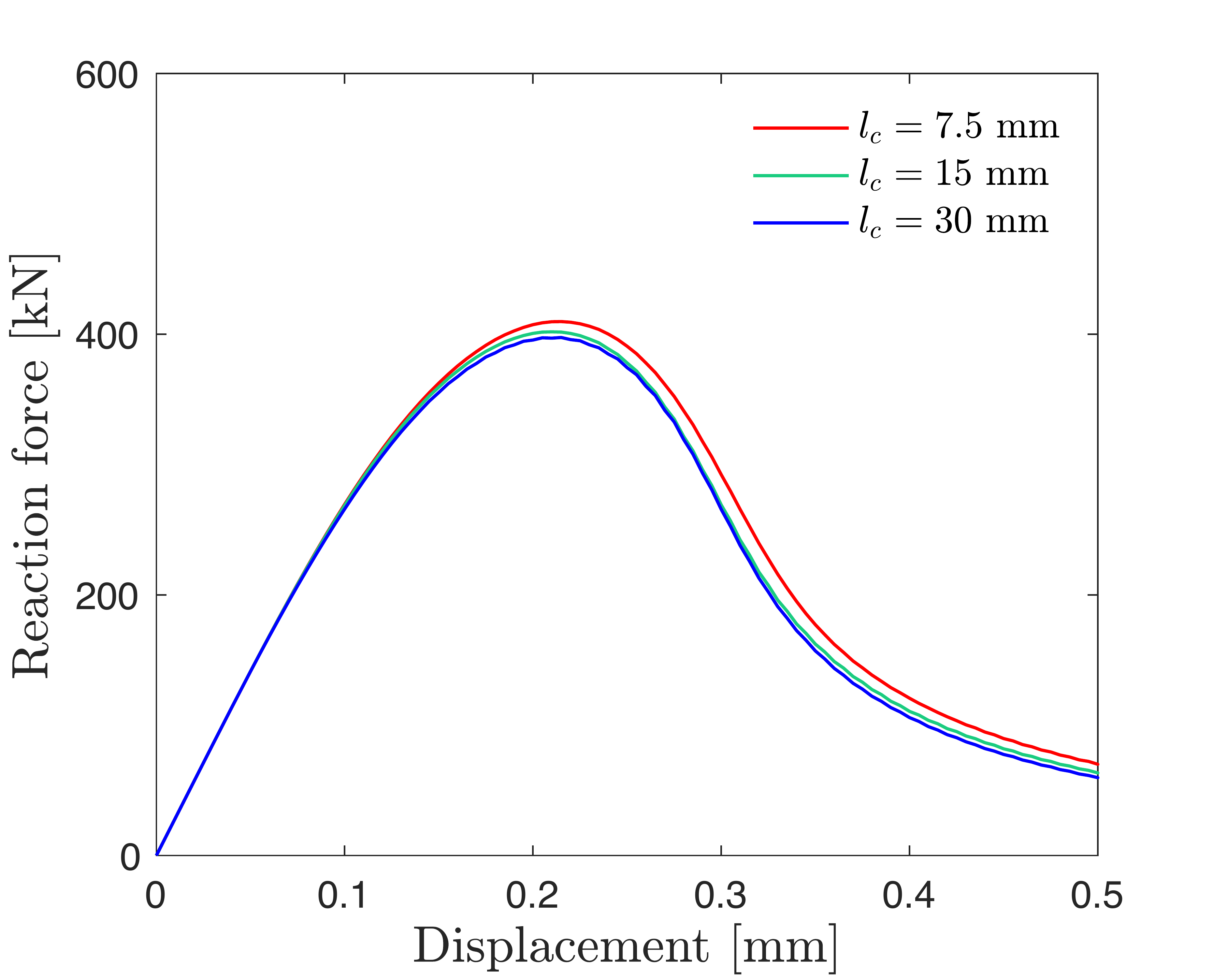}}
\caption{The force-displacement curves obtained from the trapezoid problem with different regularization lengths: non-polar case.}
\label{fig:trapezoid_response}
\end{figure}

The material is assumed to be similar to the concrete studied in \citet{Lorentz2012}. 
The material parameters for this example are chosen as follows: Young's modulus $E = 30$ GPa, Poisson's ratio $\nu = 0.2$, critical energy release rate $\mathcal{G}_c = 0.1$ N/mm, and threshold energy density $\psi_{\text{crit}} = 0.1$ kJ/m$^3$. 
We also set coupling number $N = 0$ and bending characteristic length $l_b = 0$ mm in order to avoid all the micropolar effects, and at the same time damage only the pure Boltzmann part, i.e., $\mathfrak{D} = \left\lbrace B \right\rbrace$, $\mathfrak{U} = \left\lbrace C, R \right\rbrace$. 
We prescribe $\Delta \bar{u}_2 = 0.5 \times 10^{-3}$ mm on the left boundaries, while all other boundaries are maintained traction-free during the simulations.

\begin{figure}[h]
\centering
\includegraphics[height=0.375\textwidth]{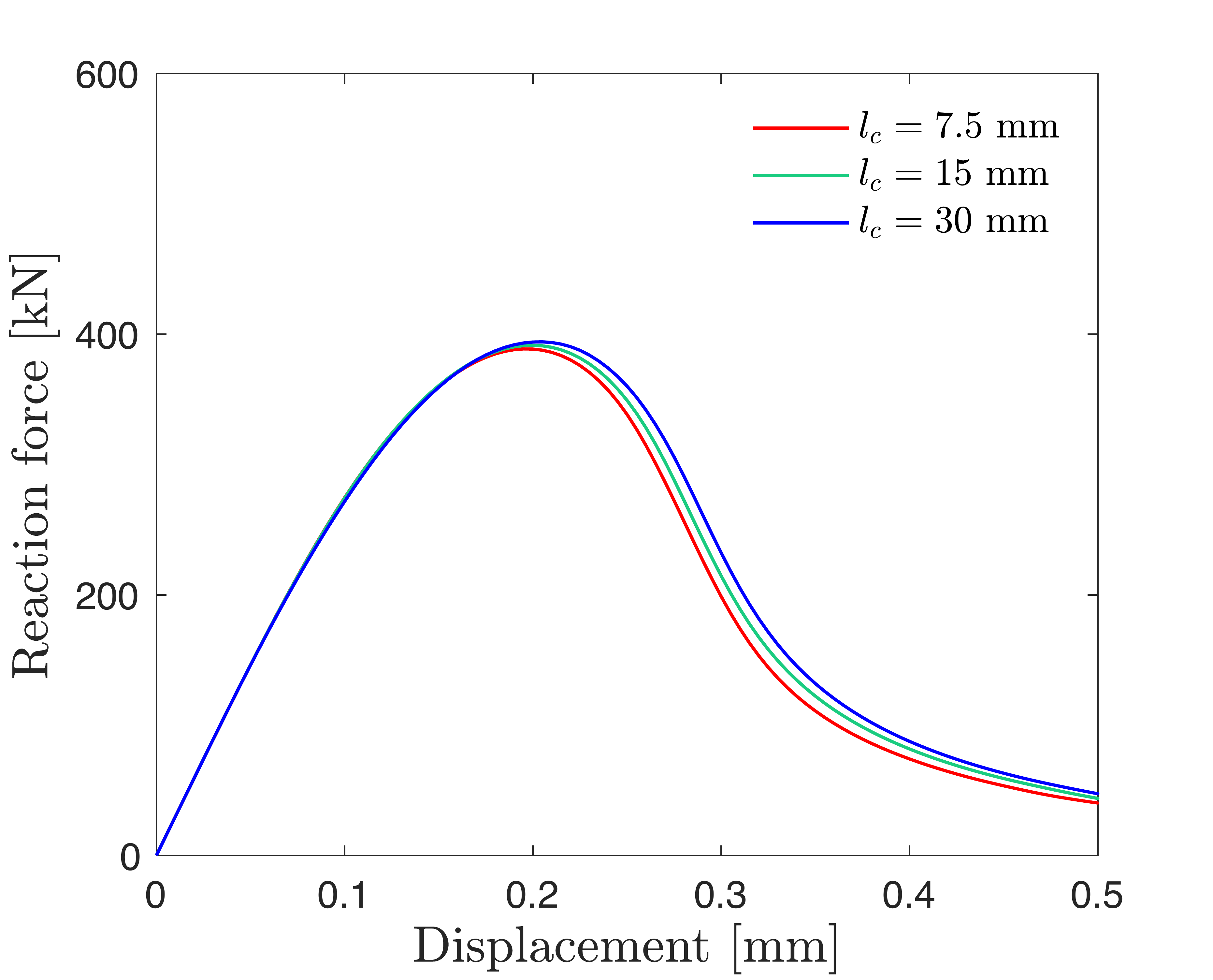}
\caption{The force-displacement curve obtained from the trapezoid problem with different regularization lengths: micropolar case, $p =10$, $l_b = 50$ mm, and $N = 0.5$.}
\label{fig:trapezoid_micropolar}
\end{figure}

Fig.~\ref{fig:trapezoid_response} shows the force–displacement curves for the trapezoid problem, corroborated by other numerical observations \citep{Lorentz2012, Geelen2019}. 
The results indicate that the shape parameter $p$ is able to influence the peak force and the overall global force-displacement responses. 
In fact, the quasi-quadratic degradation function enables us to not only tailor the threshold for the elastic region by controlling $\psi_{\text{crit}}$, but also tune the peak stress by varying shape parameter $p$. 
As pointed out in \citet{Geelen2019}, higher value of $p$ tend to significantly elongate the length of the fracture process zone, such that the stresses in the process zone can effectively be smeared over a large distance. 
As a result, we can observe that an increasing value of $p$ yields the global force-displacement response where the effect of using different length scale parameters $l_{c}$ becomes negligible, as previously reported in \citet{Lorentz2012, wu2018length, Geelen2019, wu2020phase}.

We then repeat the same problem with the micropolar material, i.e., the case where $\mathfrak{D} = \left\lbrace B, C, R \right\rbrace$, $\mathfrak{U} = \emptyset$. 
Recall Section ~\ref{1d_analysis} that the regularization length insensitive response is expected for the micropolar material as well. 
For this problem, we set coupling number $N = 0.5$, and bending characteristic length $l_b = 50$ mm, while we choose the shape parameter as $p = 10$ which produced regularization length insensitive results in Fig.~\ref{fig:trapezoid_p100}. 
As illustrated in Fig.~\ref{fig:trapezoid_micropolar}, the force-displacement curve confirms that our choice $p=10$ yields the regularization length insensitive global response for micropolar material as well. 
From this numerical example, we again spotlight the fact that the high value of the shape parameter $p$ yields the regularization length insensitive response in both non-polar and micropolar material.

\subsection{Single edge notched tests}
\label{single_notch}
\begin{figure}[h]
\centering
\includegraphics[width=0.425\textwidth]{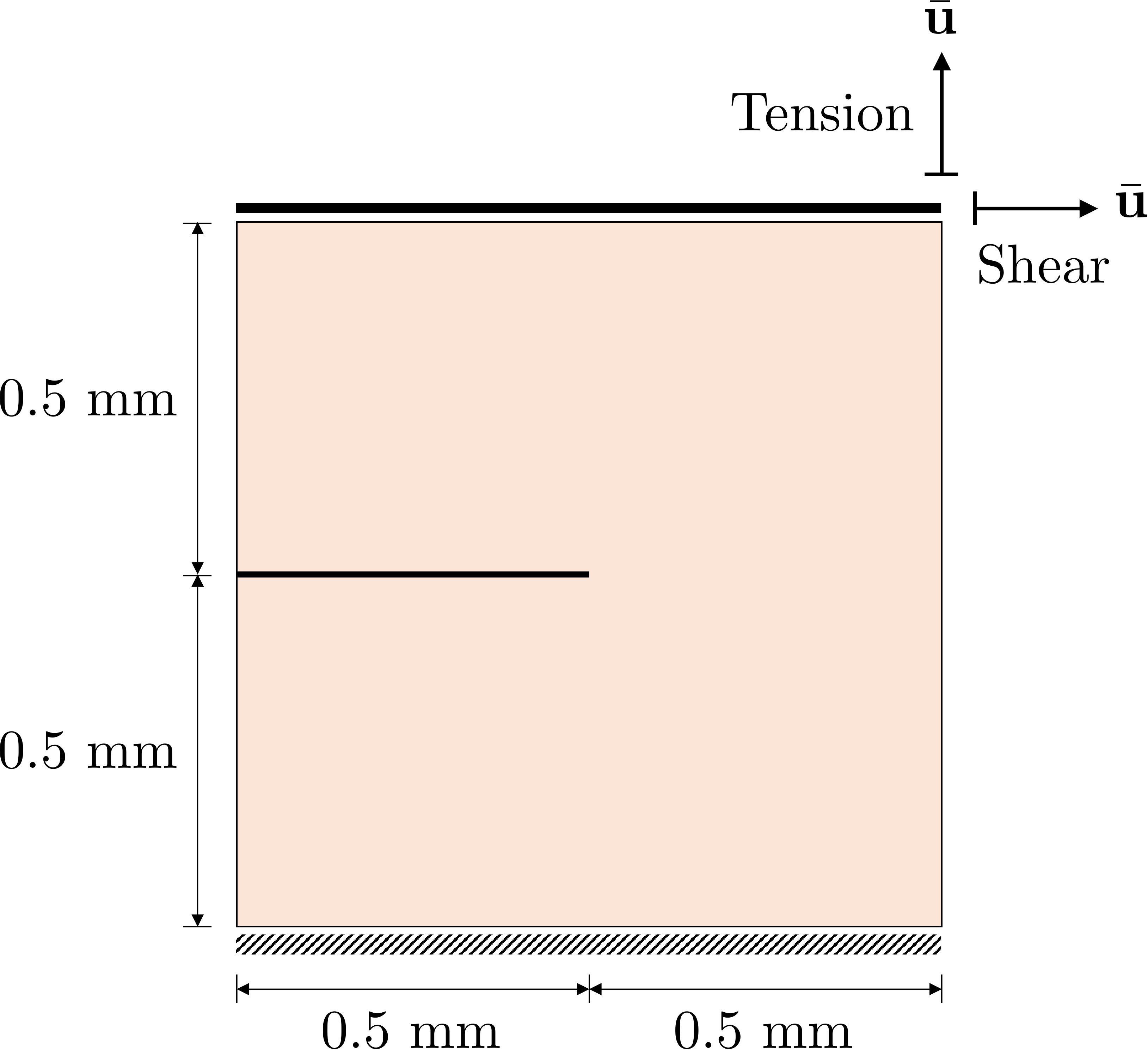}
\caption{Schematic of geometry and boundary conditions for the single edge notched tests.}
\label{fig:single_edge_notch_domain}
\end{figure}

We now consider the classical boundary value problem, which serves as a platform to investigate the size effect of elasticity and energy dissipation on the crack nucleation and propagation. 
The problem domain is a square plate that has an initial horizontal edge crack placed at the middle from the left to the center (Fig.~\ref{fig:single_edge_notch_domain}). 
Similar to the previous studies \citep{Miehe2010, Borden2012}, we choose material parameters as: $E = 210$ GPa, $\nu = 0.3$, $\mathcal{G}_c = 2.7$ N/mm, $l_c = 0.008$ mm, and $\psi_{\text{crit}} = 10$ MJ/m$^3$. 
Numerical experiments are conducted with different bending characteristic lengths: $l_b =$ 0.0, 0.01, 0.05, and 0.25 mm while the coupling number is held fixed as $N = 0.5$. Also, for this problem we damage all the energy density parts, i.e., $\mathfrak{D} = \left\lbrace B, C, R \right\rbrace$, $\mathfrak{U} = \emptyset$. 
As illustrated in Fig.~\ref{fig:single_edge_notch_domain}, two different types of simulations are conducted with the same specimen: the pure tensile test with prescribed vertical displacement $\Delta \bar{u}_2 = 2.0 \times 10^{-5}$ mm; and the pure shear test with prescribed horizontal displacement $\Delta \bar{u}_1 = 2.0 \times 10^{-5}$ mm. 
In both cases, the displacements are prescribed along the entire top boundary, while the bottom part of the domain is fixed. 

\begin{figure}[h]
\centering
\includegraphics[width=0.9\textwidth]{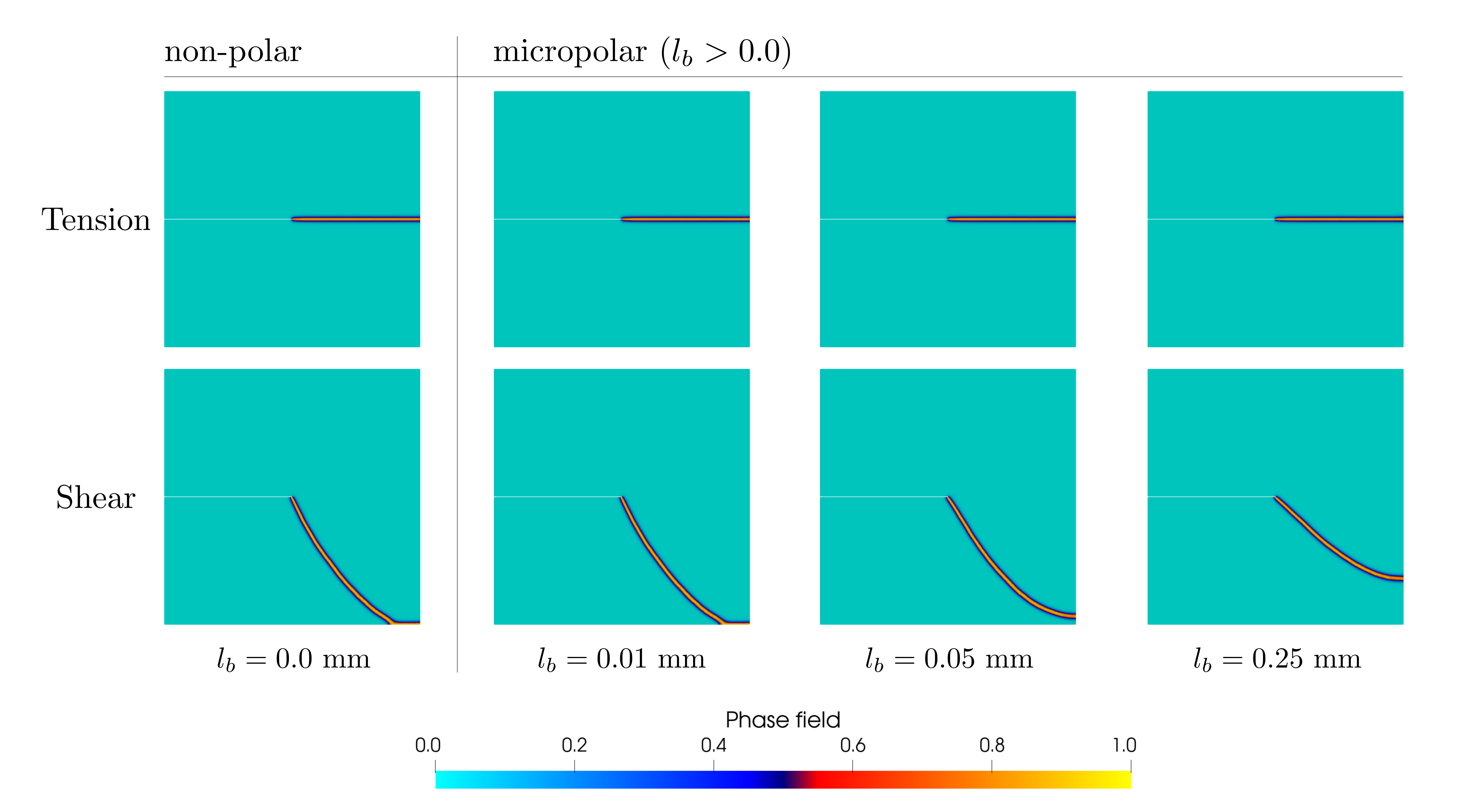}
\caption{Fracture patterns for single edge notched tests under different micropolar characteristic length $l_b$.}
\label{fig:single_edge_notch_crackpath}
\end{figure}

Fig.~\ref{fig:single_edge_notch_crackpath} illustrates the crack trajectories at the completely damaged stages for both tension and shear tests with different the bending characteristic lengths. 
The results clearly show that pure tensile loading exhibits the same crack pattern regardless of $l_b$.
This result is expected, as introducing the micro-polar effect should not break the symmetry of the boundary value problems in pure Mode I loading. 
Interestingly, the higher-order constitutive responses have a profound impact on the crack propagation direction in the Mode II simulations. 
As shown in Fig.~\ref{fig:single_edge_notch_crackpath}, the micropolar effect leads to a propagation direction bends counterclockwise. 
As the driving force for the phase field is affected by the micro-rotation induced by the coupling between shear and micro-rotation, this affect the energy dissipation mechanism and ultimately the energy minimizer of the action functional that provides the deformed configuration and crack patterns. 

As pointed out in \citet{yavari2002fractal}, the particles near crack tip resist micro-rotation and separation (i.e., interlocking), and the crack propagation mechanism in micropolar continuum therefore consists of the following steps. 
First, micro-rotational bonding between adjacent particles at the crack tip breaks and the particles starts to rotate with respect to each other. 
Second, the particles then move apart and the adjacent set of particles become the next crack-tip particles. 
Based on the mechanism, the crack path that minimizes the effort on breaking the micro-rotational bond (i.e., the path that maximizes the dissipation) is equivalent to the shortest path towards the boundary if the material is isotropic and homogeneous (e.g., a horizontal crack growth from the notch tip for the pure tensile loading case). 
The observed crack patterns shown in Fig.~\ref{fig:single_edge_notch_crackpath} is consistent with this interpretation.  Furthermore, Fig.~\ref{fig:single_edge_notch_crackpath} also provides evidence to support that the non-polar model is a special case of the micropolar model in which $l_{b} \approx 0$. 
With a sufficiently small bending characteristic length $l_{b}$, the difference in crack patterns for the non-polar and micropolar cases are negligible. 
The coupling and the micro-rotational parts ($\psi_e^C$ and $\psi_e^R$) then become significant enough to play a bigger role for crack growth as the bending characteristic length $l_{b}$ increases. 

\begin{figure}[h]
\centering
\includegraphics[width=0.95\textwidth]{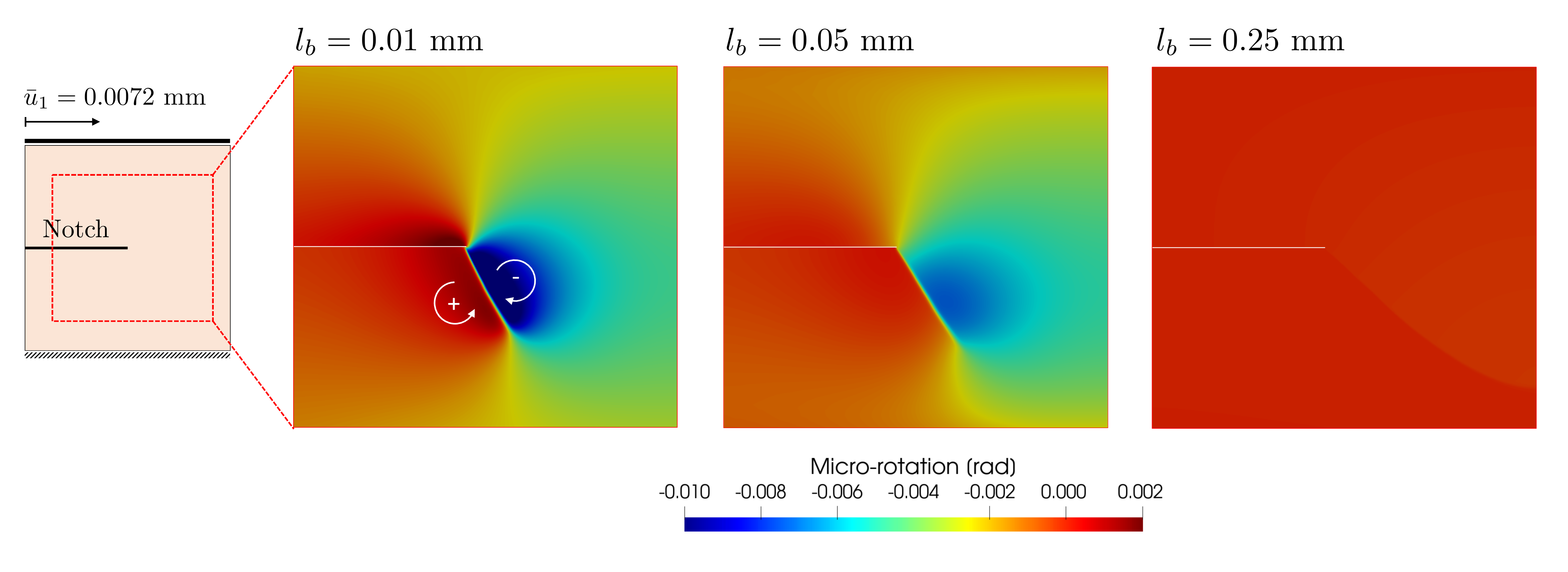}
\caption{Resultant micro-rotation field for single edge notched shear tests where $\bar{u}_1 = 0.0072$ mm.}
\label{fig:single_edge_notch_microrotation}
\end{figure}

As illustrated in Fig.~\ref{fig:single_edge_notch_microrotation}, separated particles tend to rotate in opposite directions since they are no longer interlocked after crack formation. 
By revisiting Eq.~\eqref{eq:engineering_material_constants} and Eq.~\eqref{eq:couple_stress}, notice that the material constant $\gamma$ that relates the fictitious undamaged couple stress $\bar{\tensor{m}}^R$ to the micro-curvature $\bar{\tensor{\kappa}}$ is proportional to the square of the characteristic lengths. 
The relationship implies that larger characteristic length leads to higher rigidity of the micropolar material, so that the separated particles tend to experience greater micro-rotation with smaller bending characteristic length.

\begin{figure}[h]
\centering
\subfigure[Global response from the tension test.]{\label{fig:single_edge_notch_mode1_response}\includegraphics[height=0.375\textwidth]{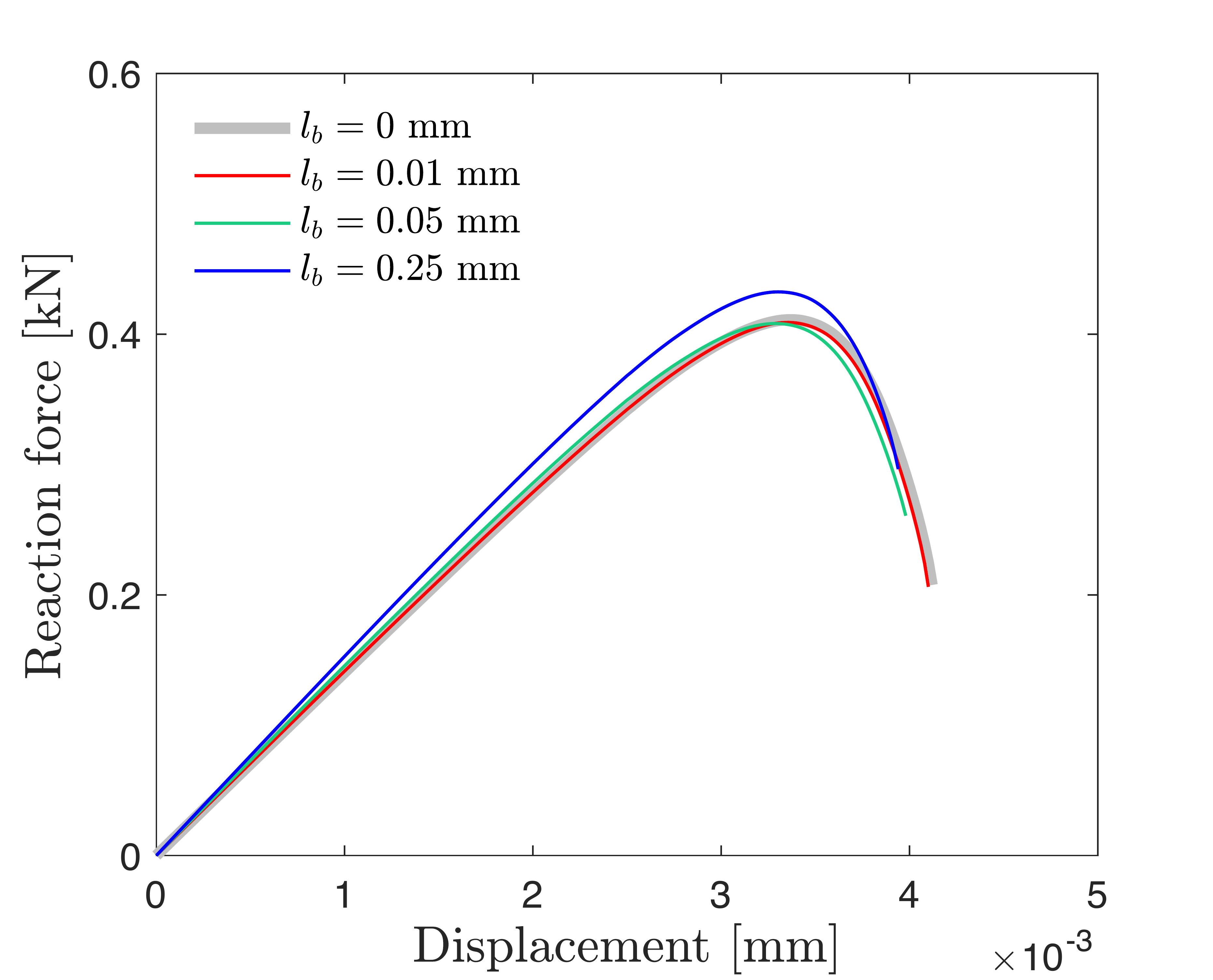}}
\hspace{0.01\textwidth}
\subfigure[Global response from the shear test.]{\label{fig:single_edge_notch_mode2_response}\includegraphics[height=0.375\textwidth]{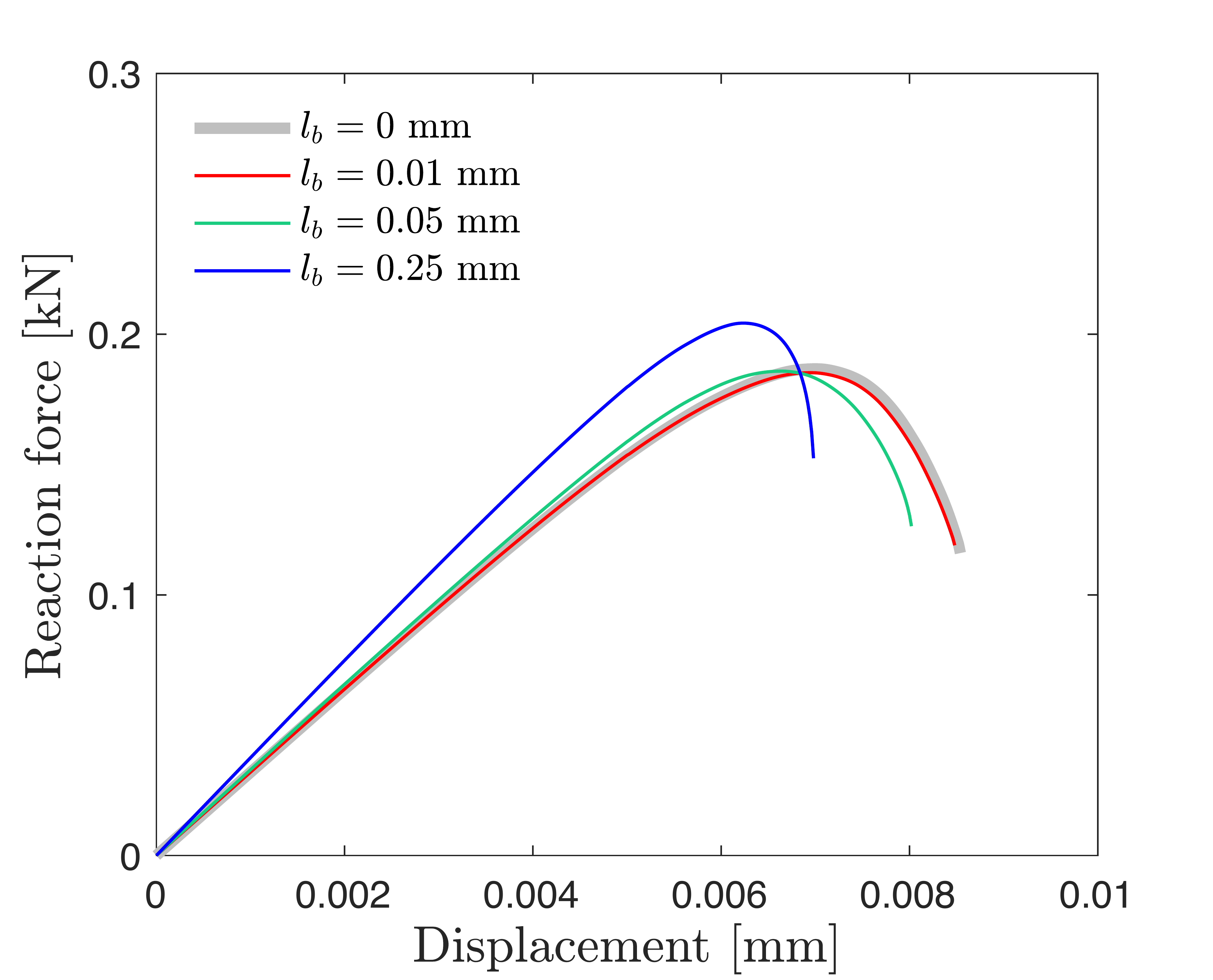}}
\caption{The force-displacement curves from the single edge notched tests with regularization length $l_c = 0.008$ mm.}
\label{fig:single_edge_notch_response}
\end{figure}

Fig.~\ref{fig:single_edge_notch_response} shows the load-deflection curves obtained from both tension and shear tests. 
The colored curves indicate the results with nonzero $l_b$, whereas the transparent gray curves denote the non-polar case.  
Since the force stress can be decomposed into two parts, e.g., $\bar{\tensor{\sigma}} = \bar{\tensor{\sigma}}^B + \bar{\tensor{\sigma}}^C$, micropolar material that possesses a large characteristic length tend to exhibit stiffer response compared to those with smaller characteristic lengths, due to the micro-continuum coupling effect. 
Unlike the tension test results in Fig.~\ref{fig:single_edge_notch_mode1_response}, the reaction forces reach their peak values under different strain level from the shearing tests [Fig.~\ref{fig:single_edge_notch_mode2_response}]. 
This again highlights that the micropolar bending characteristic length affects the crack pattern, which in turn reflects different global response for the same boundary value problem. 

\subsection{Asymmetric notched three-point bending tests}
\label{three_point_bending}
\begin{figure}[h]
\centering
\includegraphics[width=0.8\textwidth]{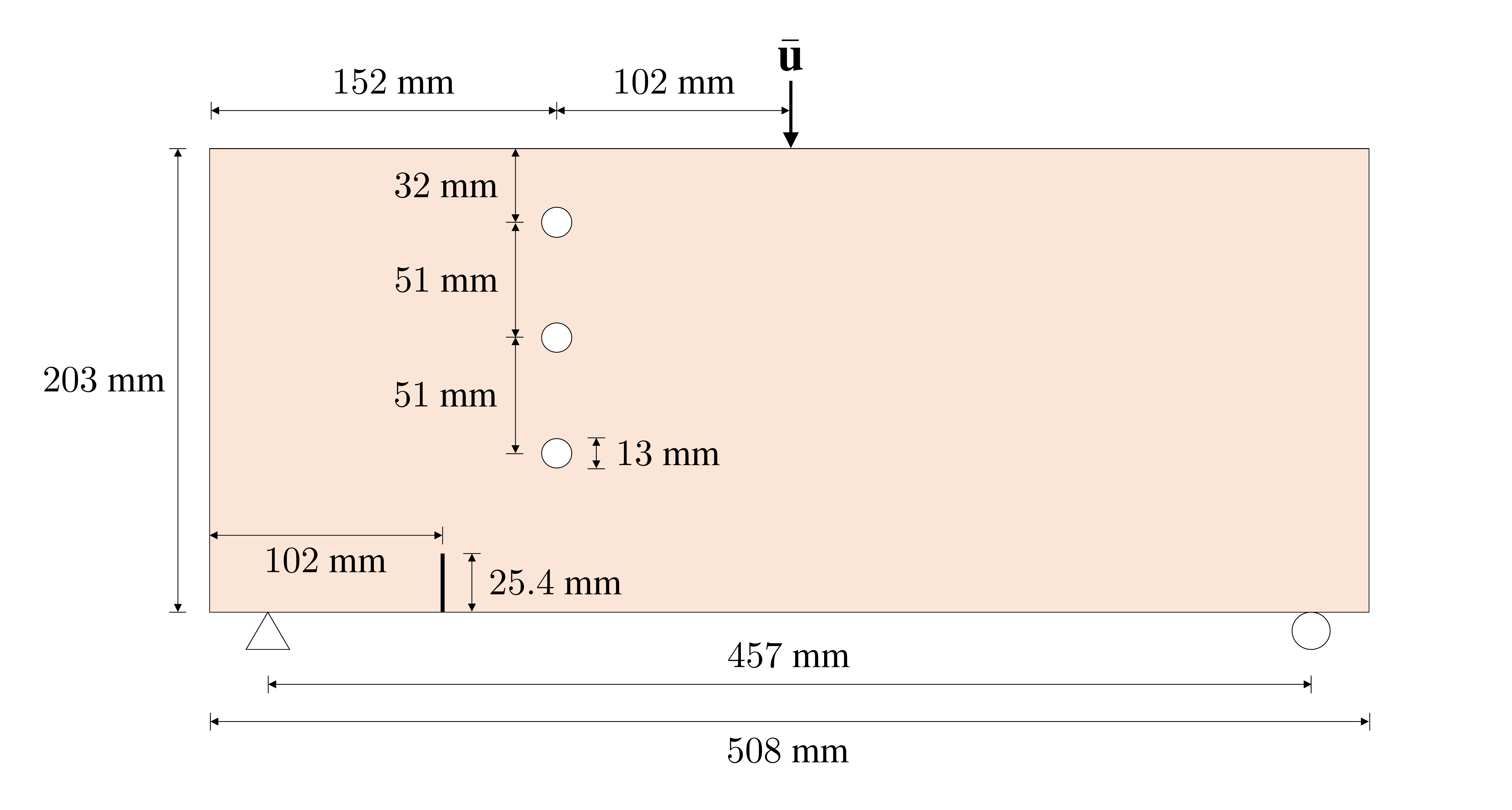}
\caption{Schematic of geometry and boundary conditions for the three-point bending tests.}
\label{fig:beam_domain}
\end{figure}

This section examines a problem originally designed by \citet{ingraffea1990probabilistic}, which involves a three-point bending of a specimen with three holes. 
The domain of the problem is illustrated in Fig.~\ref{fig:beam_domain}. 
Since previous studies \citep{ingraffea1990probabilistic, Miehe2010, patil2018adaptive, qinami2019circumventing} have shown that different crack patterns can be observed depending on the notch depth and its position, we only focus on the case where the notch depth is set to be 25.4 mm.

We consider a specimen composed of a micropolar material and choose Boltzmann material parameters close to the properties of Plexiglas specimen tested by \citet{ingraffea1990probabilistic}: $E = 3.2$ GPa, $\nu = 0.3$, $\mathcal{G}_c = 0.31$ N/mm, $l_c = 1.0$ mm, and $\psi_{\text{crit}} = 7.5$ kJ/m$^3$. 
For this problem, we assume that all the energy density parts can be degraded, i.e., $\mathfrak{D} = \left\lbrace B, C, R \right\rbrace$, $\mathfrak{U} = \emptyset$.

Within the problem domain (Fig.~\ref{fig:beam_domain}), we first attempt to investigate the effect of coupling number on the crack trajectory by conducting multiple numerical tests with different values: $N = 0.1$, $N = 0.5$, and $N=0.9$, while bending characteristic length is held fixed as $l_b = 10.0$ mm. 
The numerical simulation conducted under a displacement-controlled regime where we keep the load increment as $\Delta \bar{u}_2 = -2.0 \times 10^{-4}$ mm.

\begin{figure}[h]
\centering
\includegraphics[width=0.725\textwidth]{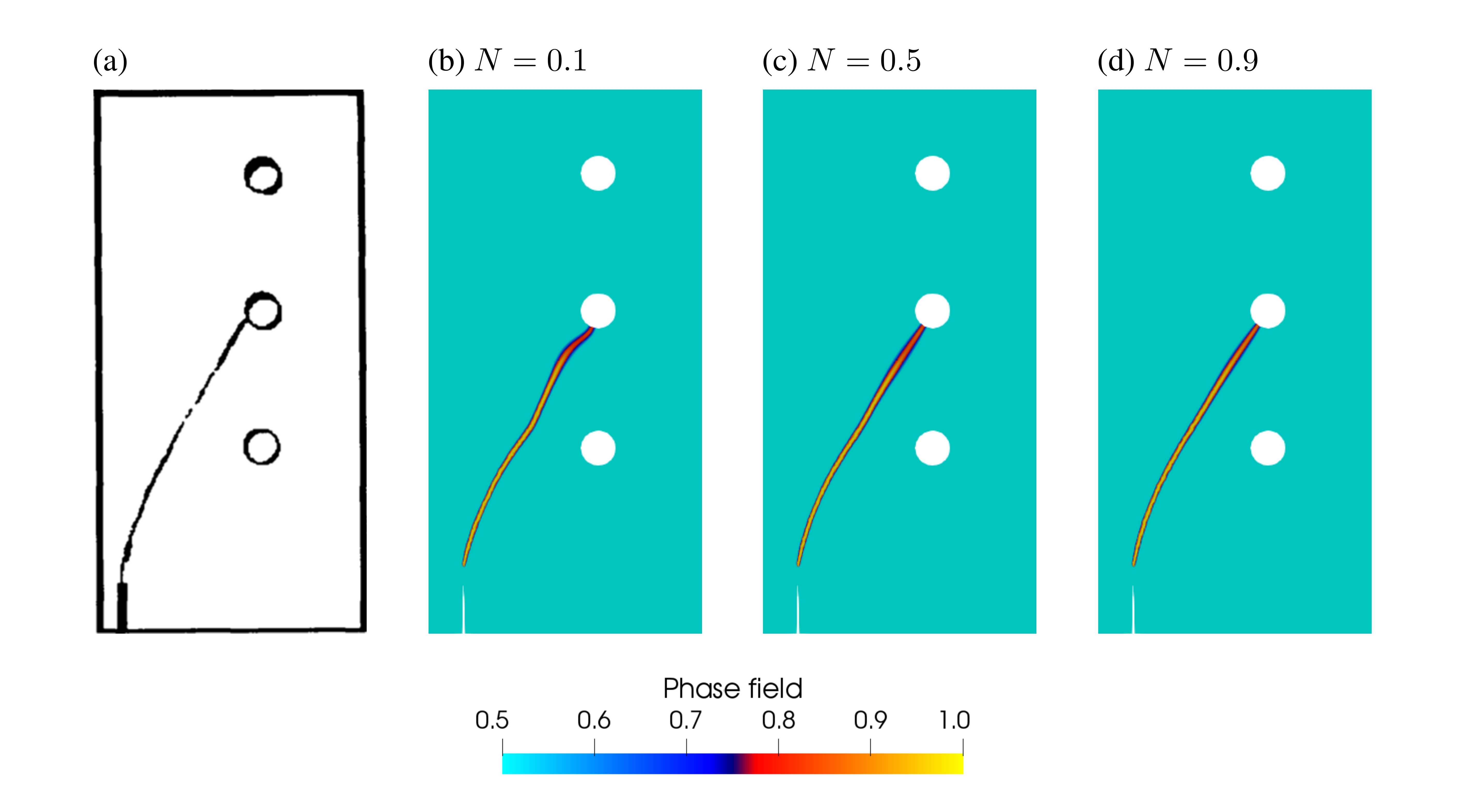}
\caption{Crack topologies for asymmetric notched three-point bending test: (a) experimentally obtained pattern by \citet{ingraffea1990probabilistic}; (b)-(d) numerically obtained parttern with different coupling number $N$.}
\label{fig:beam_N}
\end{figure}

\begin{figure}[h]
\centering
\includegraphics[height=0.375\textwidth]{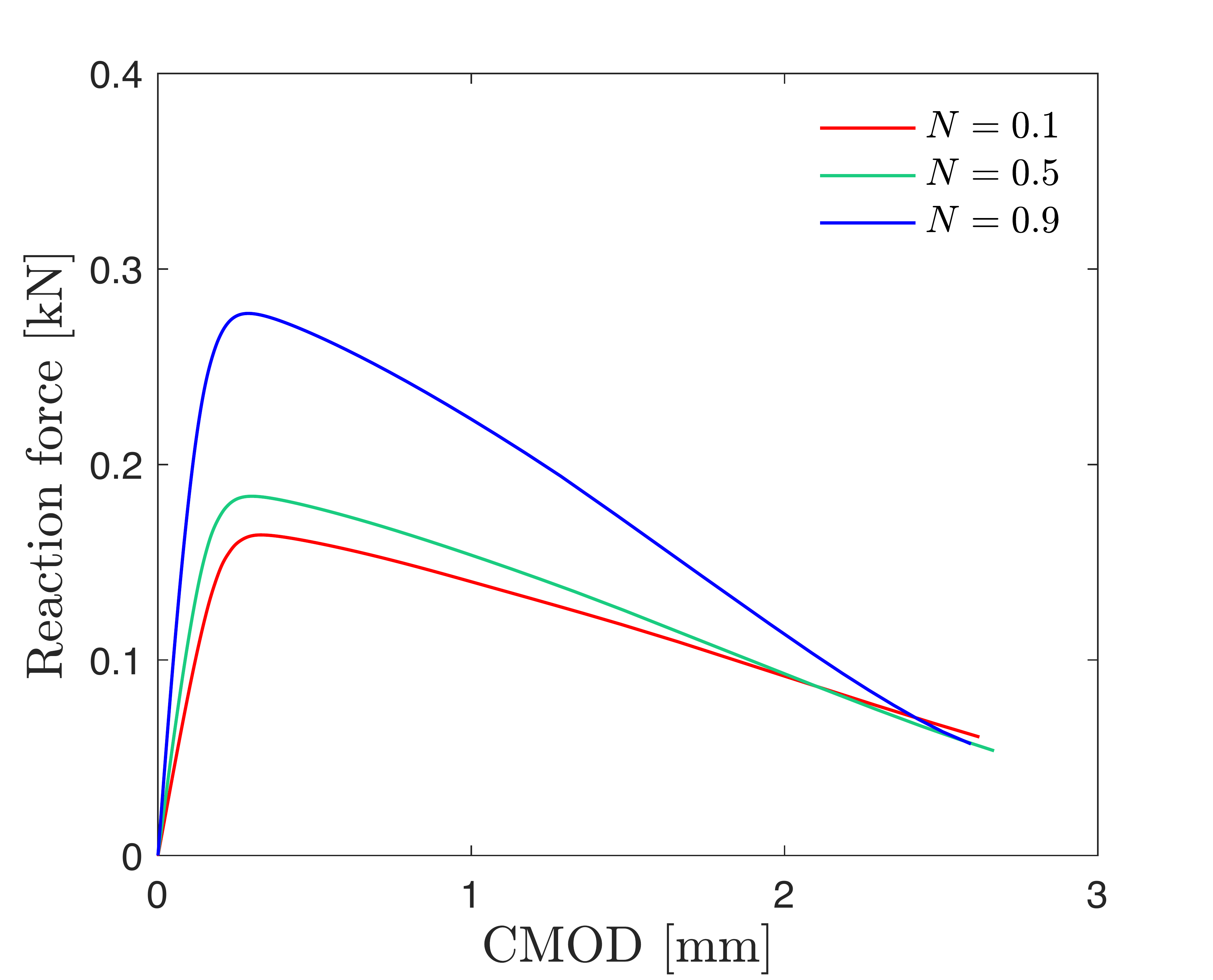}
\caption{Load-CMOD curves from the three-point bending test.}
\label{fig:beam_force_disp_N}
\end{figure}

Fig.~\ref{fig:beam_N} illustrates the crack trajectories obtained by numerical experiments with different coupling number $N$ in comparison to the experimental result \citep{ingraffea1990probabilistic}, while Fig.~\ref{fig:beam_force_disp_N} shows the measured reaction forces as a function of crack-mouth-opening-displacement (CMOD). 
Similar to the experimental results in Fig.~\ref{fig:beam_N}(a), numerical results [Fig.~\ref{fig:beam_N}(b)-(d)] show that the cracks tend to deflect towards the holes, eventually coalescing with the intermediate one. 
However, taking a closer look at Fig.~\ref{fig:beam_N}(b)-(d), one can observe the slight differences on the concavity of the crack topologies especially when the crack passes close to the bottom hole. 
As we highlighted in Section \ref{single_notch}, the crack tends to propagate through the path where the energy that takes to break the micro-rotational bond is minimized. 
At the same time, based on Saint-Venant's principle, the crack trajectory can also be affected by the bottom hole due to an increase of the singularity \citep{qinami2019circumventing}. 
In this specific platform, it can thus be interpreted that the crack propagation may result from the competition between the two, based on the obtained results. 
Since the material that possesses higher degree of micropolarity requires more energy to break the micro-rotational bond, the bottom hole effect on the crack trajectory becomes negligible as coupling number $N$ increases [Fig.~\ref{fig:beam_N}(b)-(d)]. 
In addition, Fig.~\ref{fig:beam_force_disp_N} implies that if more energy is required to break the micro-rotational bond, it results in higher material stiffness in the elastic regime, supporting our interpretation.

\begin{figure}[h]
\centering
\includegraphics[width=0.725\textwidth]{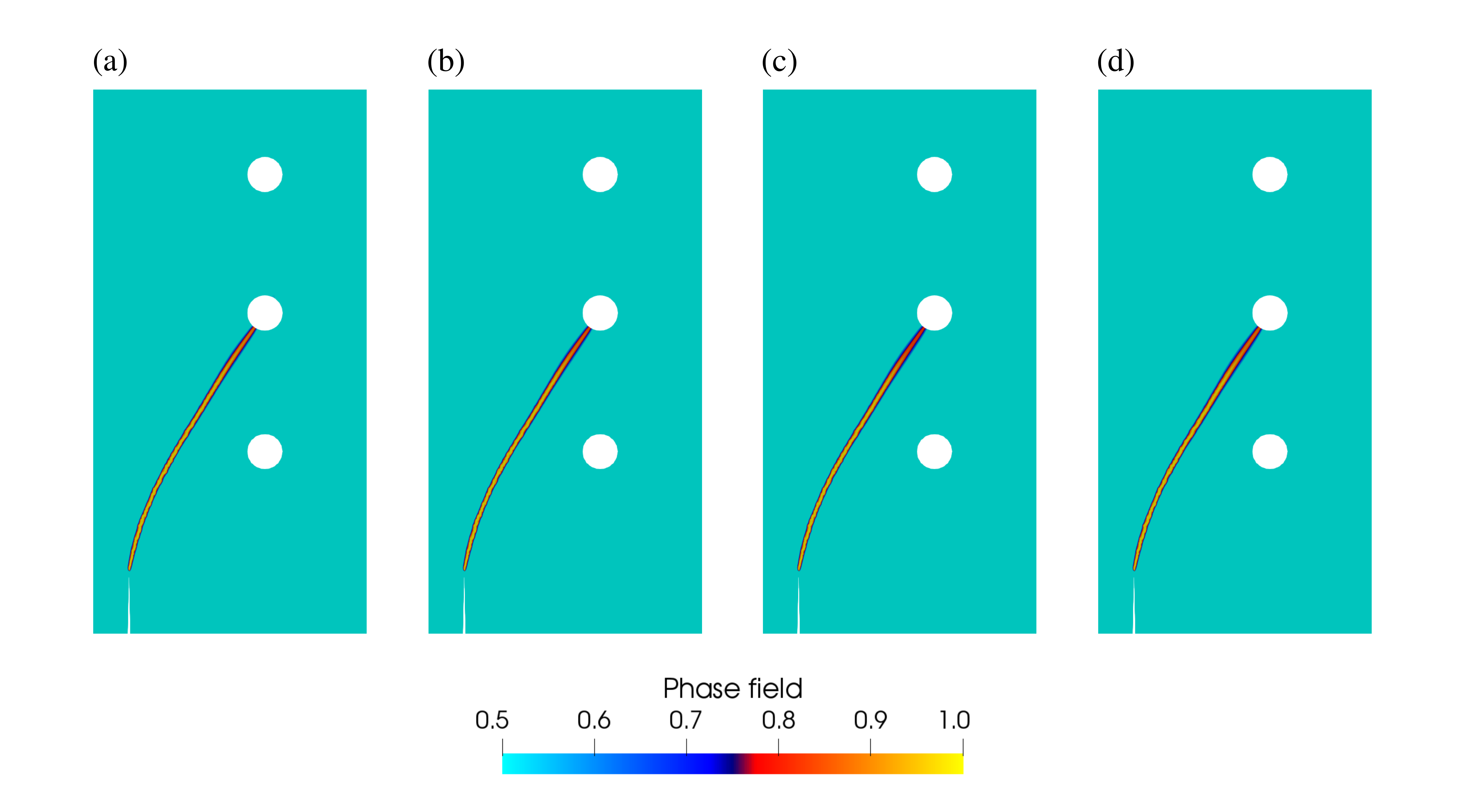}
\caption{Observed crack patterns for asymmetric notched three-point bending tests: (a) $\Delta \bar{u}_2 = -2.0 \times 10^{-4}$ mm; (b) $\Delta \bar{u}_2 = -4.0 \times 10^{-4}$ mm; (c) $\Delta \bar{u}_2 = -8.0 \times 10^{-4}$ mm; (d) $\Delta \bar{u}_2 = -16.0 \times 10^{-4}$ mm.}
\label{fig:beam_u}
\end{figure}

We then conduct a brief sensitivity analysis with respect to the time discretization (i.e., prescribed displacement increment $\Delta \bar{u}_2$) within the same problem domain, while we set the coupling number to be $N=0.9$ during the analysis. 
Fig.~\ref{fig:beam_u} illustrates the simulated crack patterns for asymmetrically notched beam with three holes, with different prescribed displacement rates, varying from $\Delta \bar{u}_2 = -4.0 \times 10^{-4}$ mm to $-16.0 \times 10^{-4}$ mm. 
Since meaningful differences in the crack trajectories are not observed, the result confirms the practical applicability of the explicit operator-splitting solution scheme, if the load increment is small enough.

\subsection{Double edge notched tests}
\label{double_notch}

\begin{figure}[h]
\centering
\includegraphics[width=0.39\textwidth]{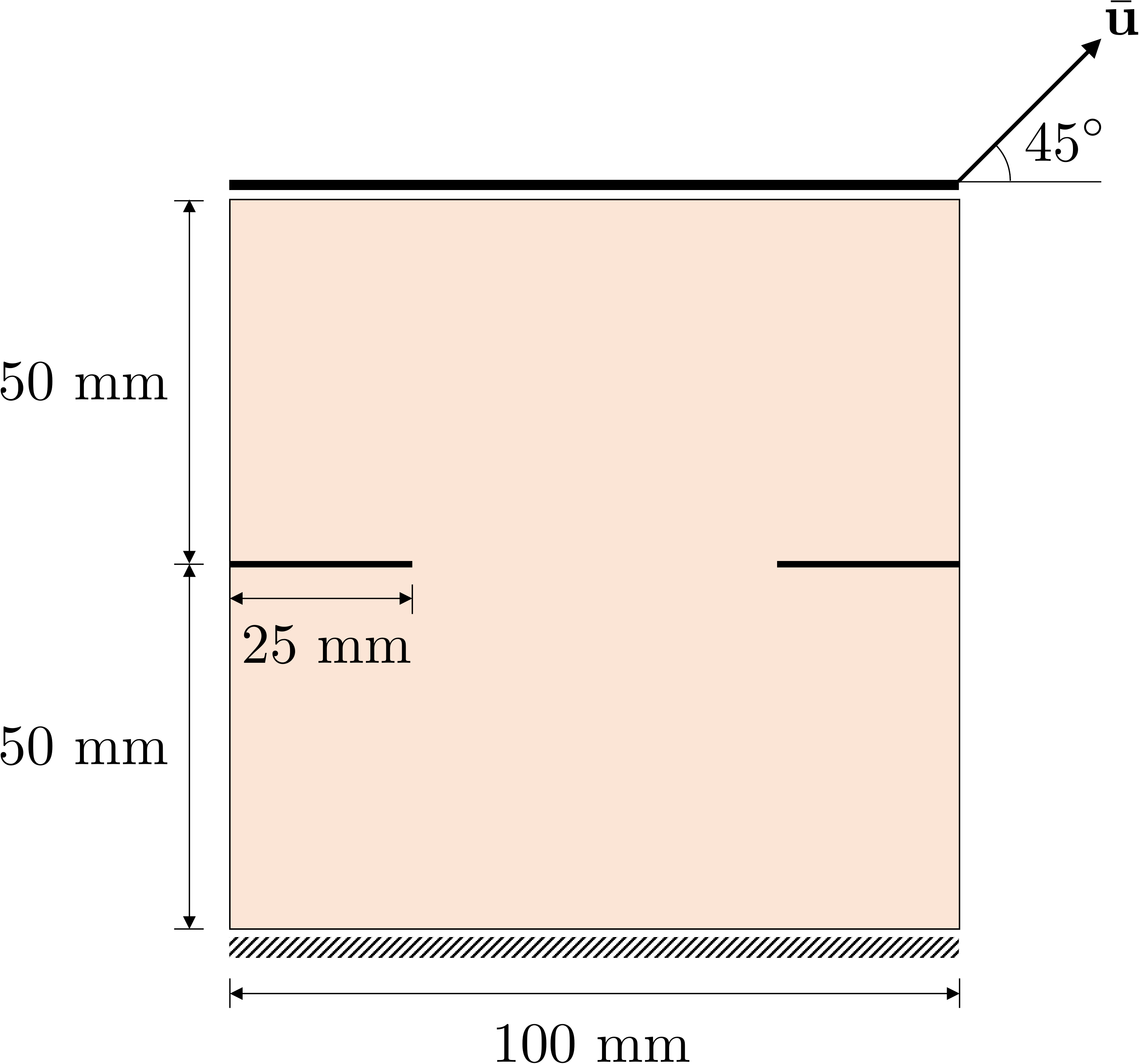}
\caption{Schematic of geometry and boundary conditions for the double edge notched tests.}
\label{fig:double_edge_notch_domain}
\end{figure}

This numerical example investigates the effect of partial degradation of the strain energy density on the crack patterns. 
As illustrated in Fig.~\ref{fig:double_edge_notch_domain}, the problem domain is a 100 mm wide and 100 mm long square plate with two 25 mm long symmetric initial horizontal edge notches at the middle. 
We assign the following material properties for this problem: $E = 30$ GPa, $\nu = 0.2$, $\mathcal{G}_c = 0.1$ N/mm, $l_c = 0.75$ mm, and $\psi_{\text{crit}} = 1.0$ kJ/m$^3$. 
Numerical experiments are simulated with bending characteristic length $l_b =$ 30.0 mm and the coupling number is set to be $N = 0.5$. 
While the bottom part of the domain is held fixed, we prescribe the displacement along the entire top boundary at an angle of 45 degrees to the horizontal direction: $\Delta \bar{u}_1 = \Delta \bar{u}_2 = 5.0 \times 10^{-4}$ mm, such that the domain is subjected to combined tensile and shear loads. 

\begin{figure}[h]
\centering
\subfigure[]{\label{fig:doublenotch_B}\includegraphics[height=0.34\textwidth]{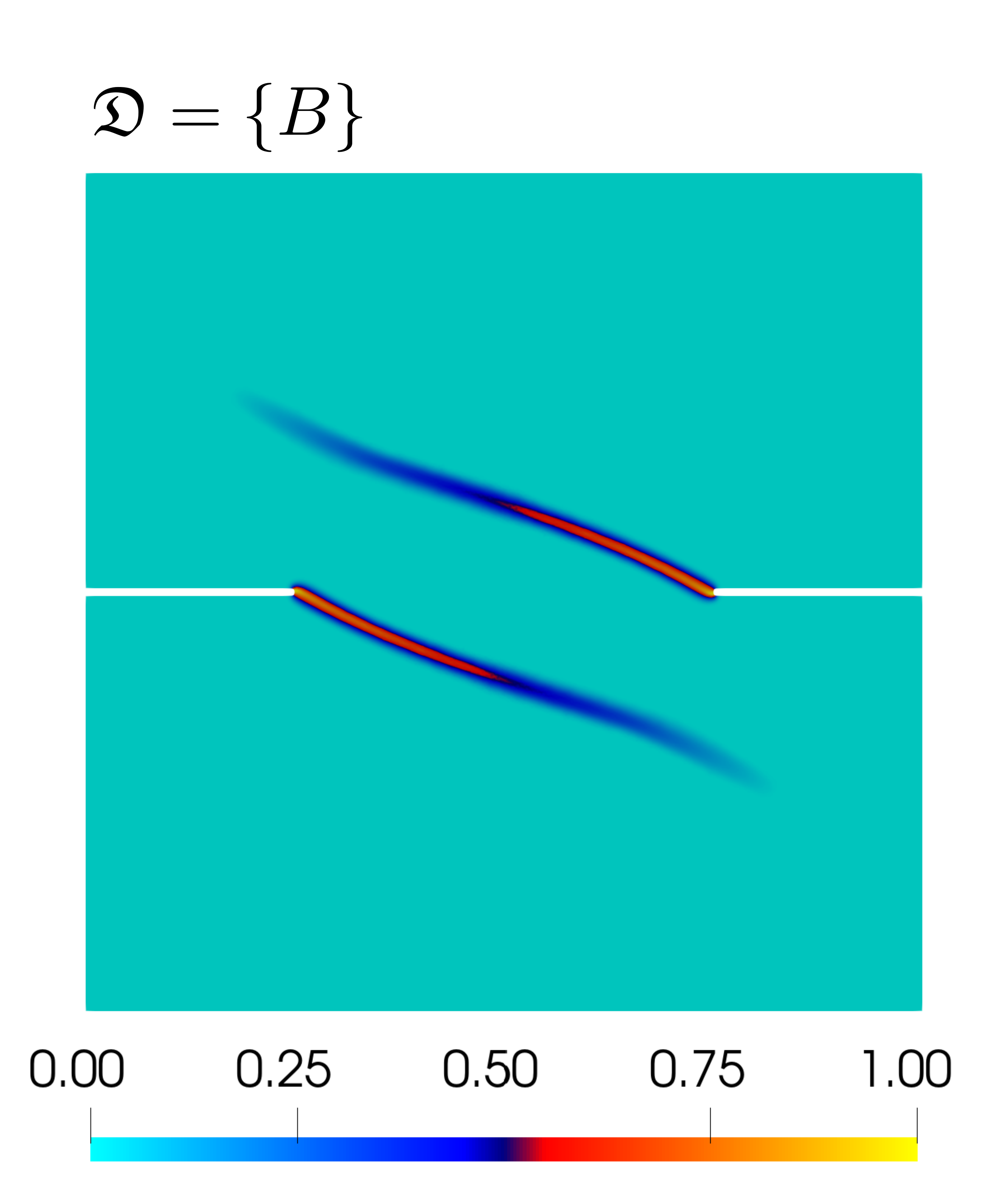}}
\hspace{0.01\textwidth}
\subfigure[]{\label{fig:doublenotch_C}\includegraphics[height=0.34\textwidth]{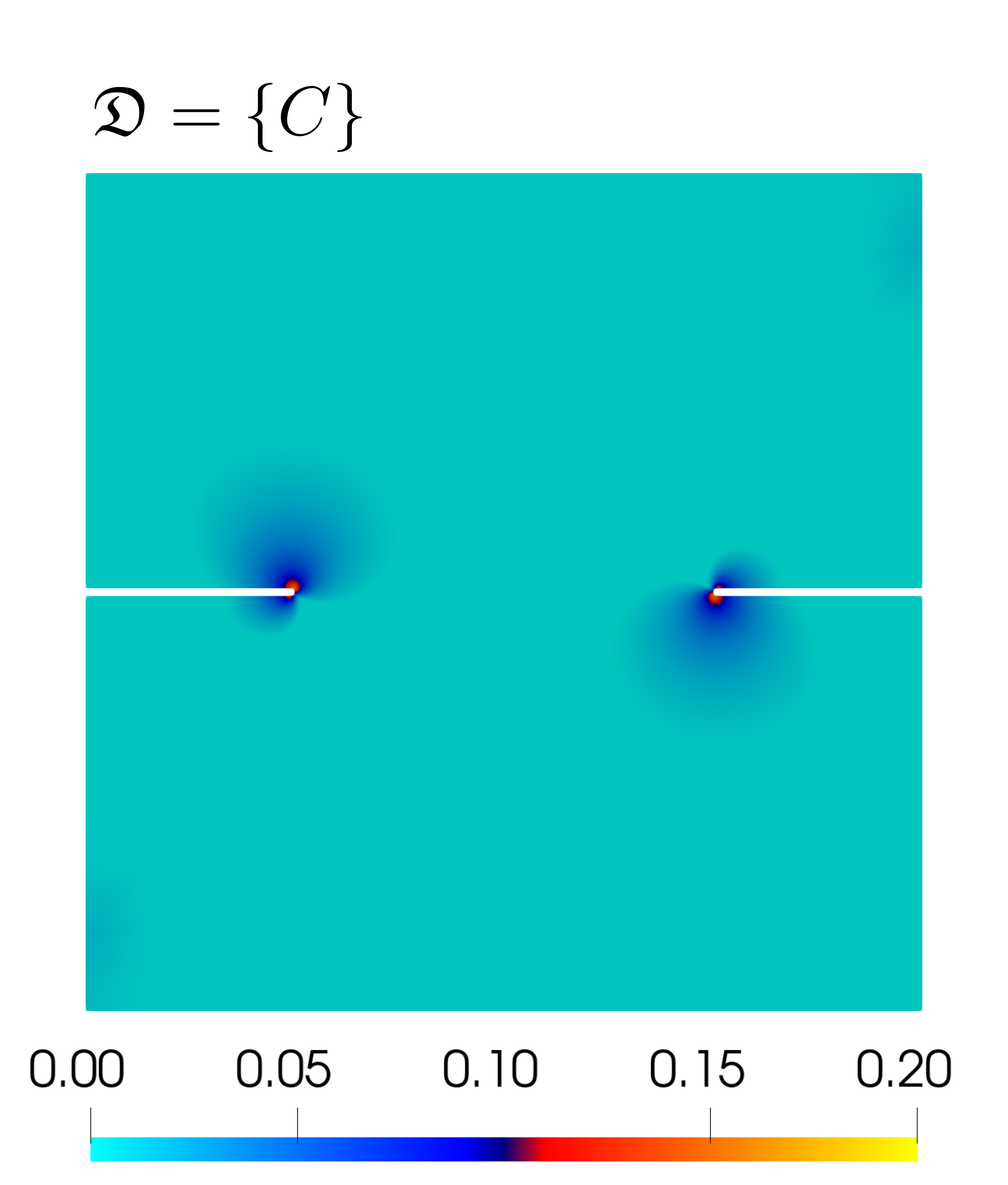}}
\hspace{0.01\textwidth}
\subfigure[]{\label{fig:doublenotch_R}\includegraphics[height=0.34\textwidth]{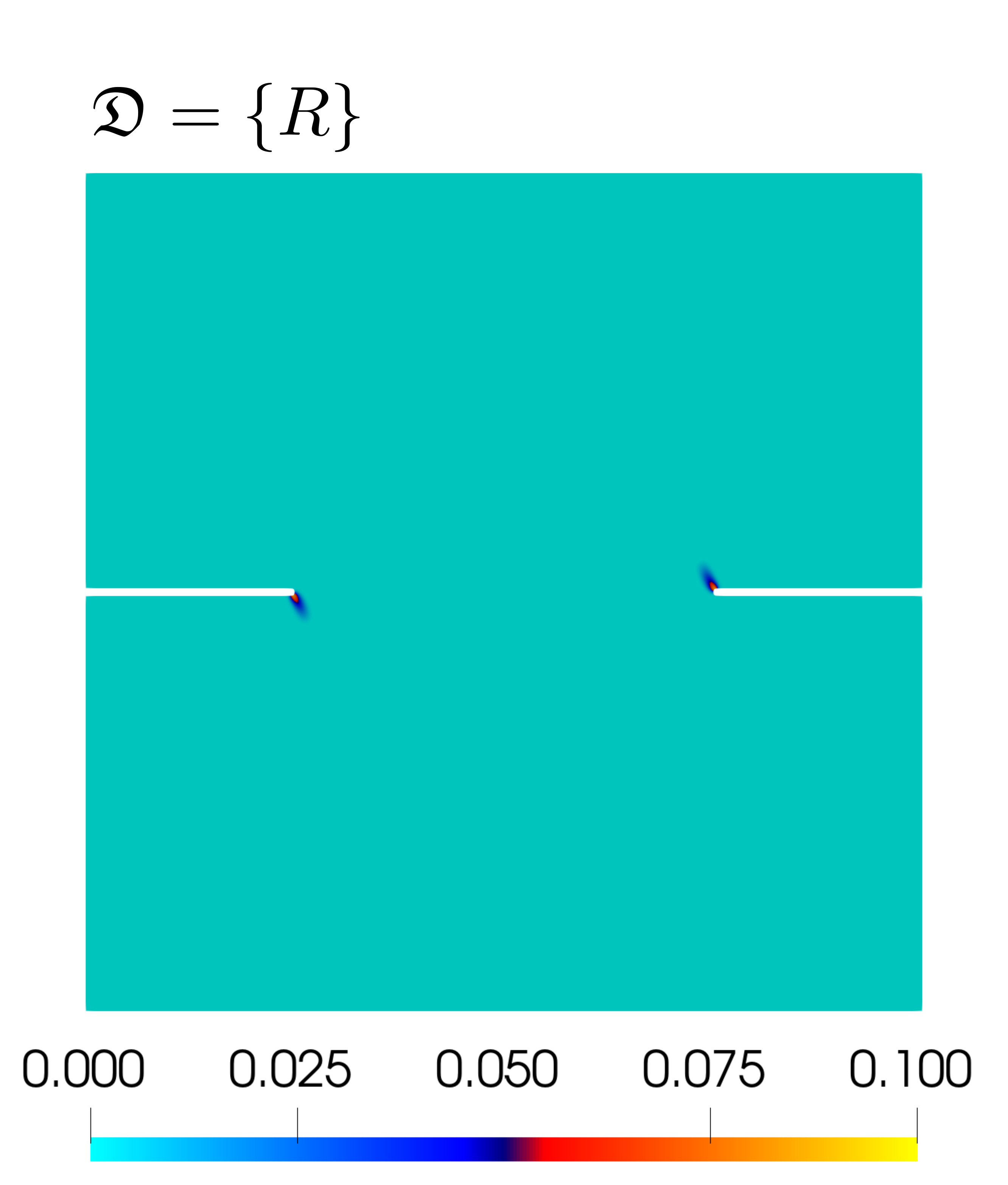}}
\caption{Crack patterns for double edge notched tests obtained by considering different degradation functions [i.e., $g_i(d) = g(d)$ if $i \in \mathfrak{D}$; $g_i(d) = 1$ otherwise] on each energy density part, where $\bar{u}_1 = \bar{u}_2 = 0.05$ mm.}
\label{fig:doublenotch_individual}
\end{figure}

\begin{figure}[h]
\centering
\subfigure[$\psi_e^B$ in kJ/m$^3$]{\label{fig:doublenotch_psiB}\includegraphics[height=0.34\textwidth]{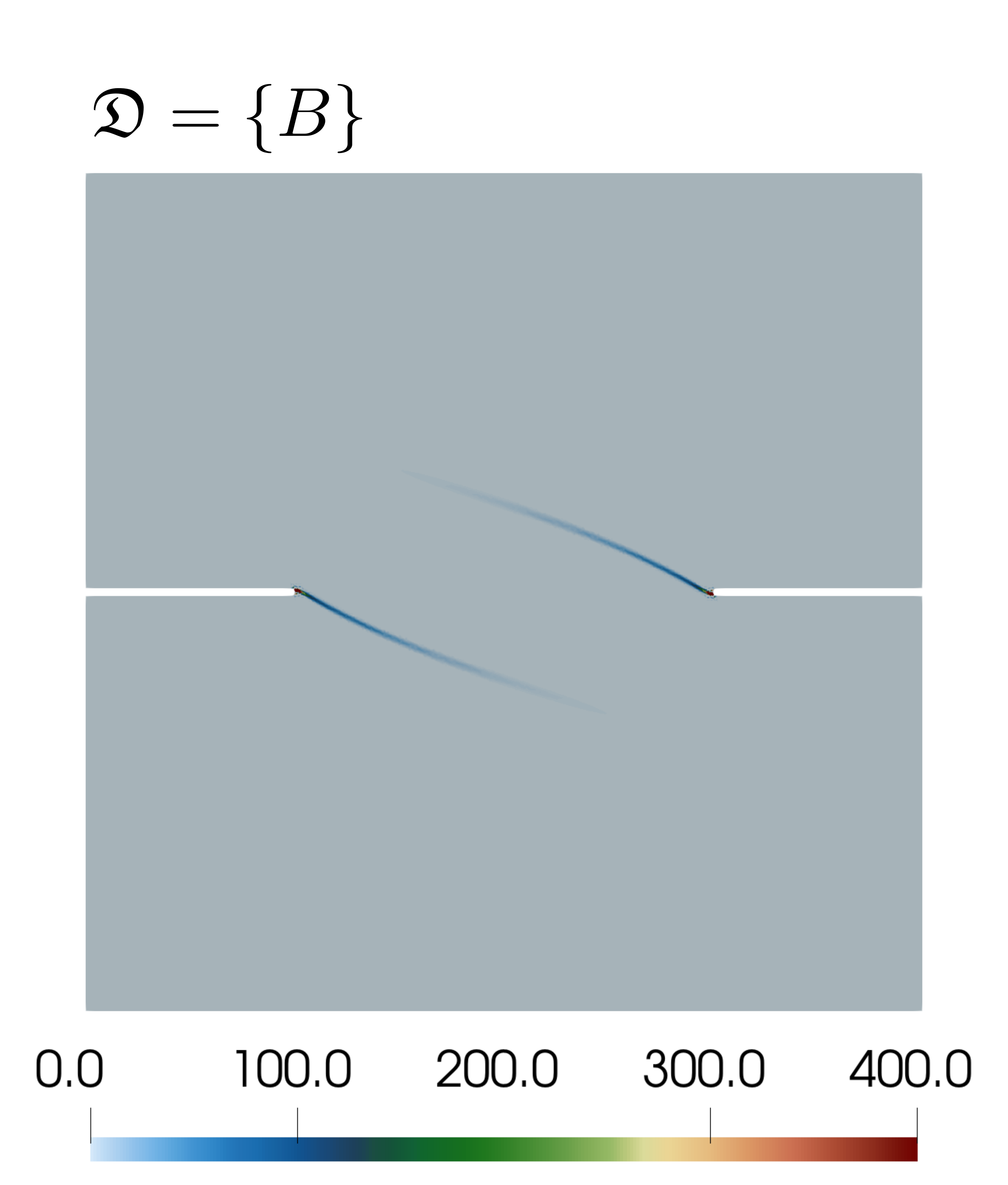}}
\hspace{0.01\textwidth}
\subfigure[$\psi_e^C$ in kJ/m$^3$]{\label{fig:doublenotch_psiC}\includegraphics[height=0.34\textwidth]{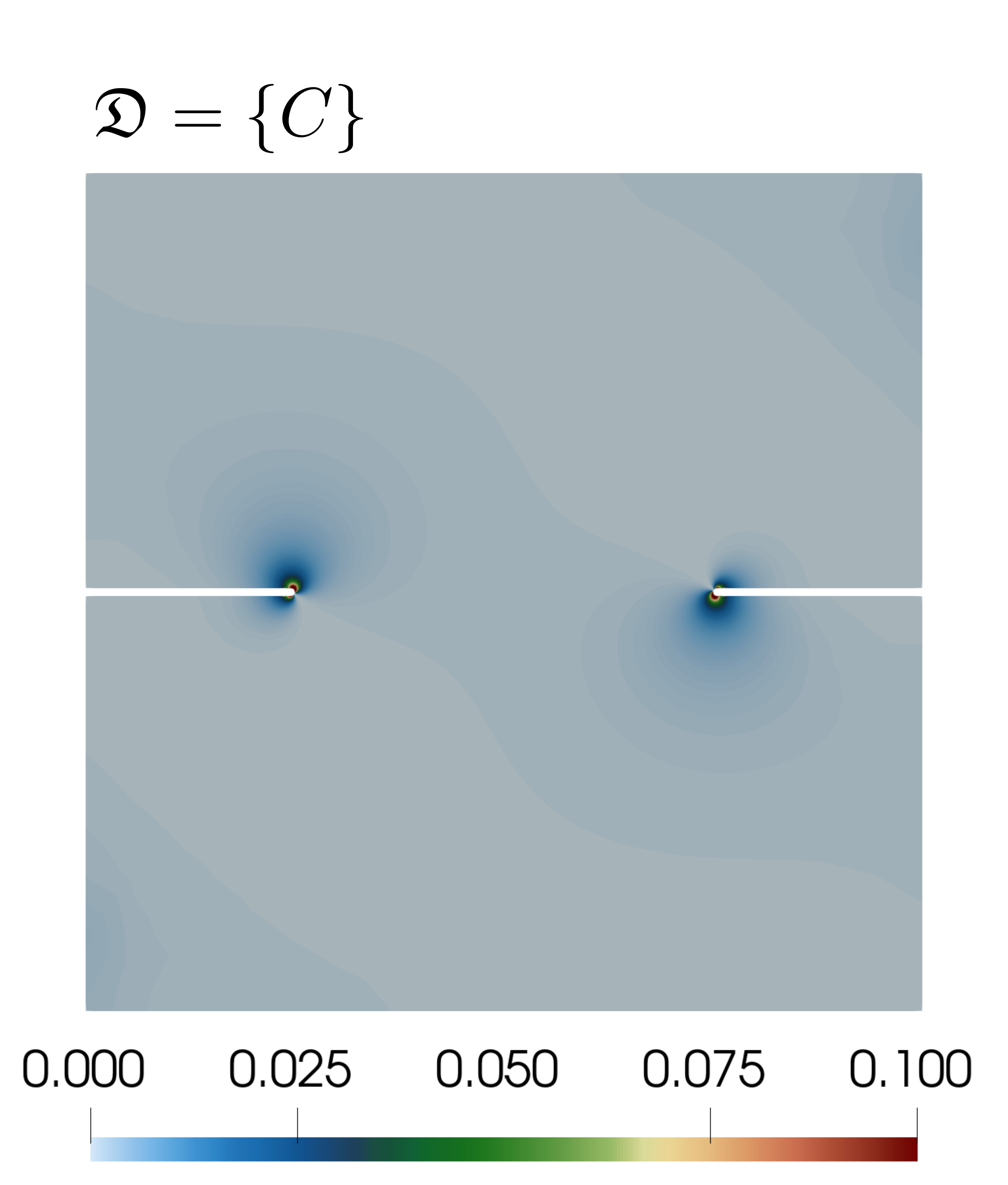}}
\hspace{0.01\textwidth}
\subfigure[$\psi_e^R$ in kJ/m$^3$]{\label{fig:doublenotch_psiR}\includegraphics[height=0.34\textwidth]{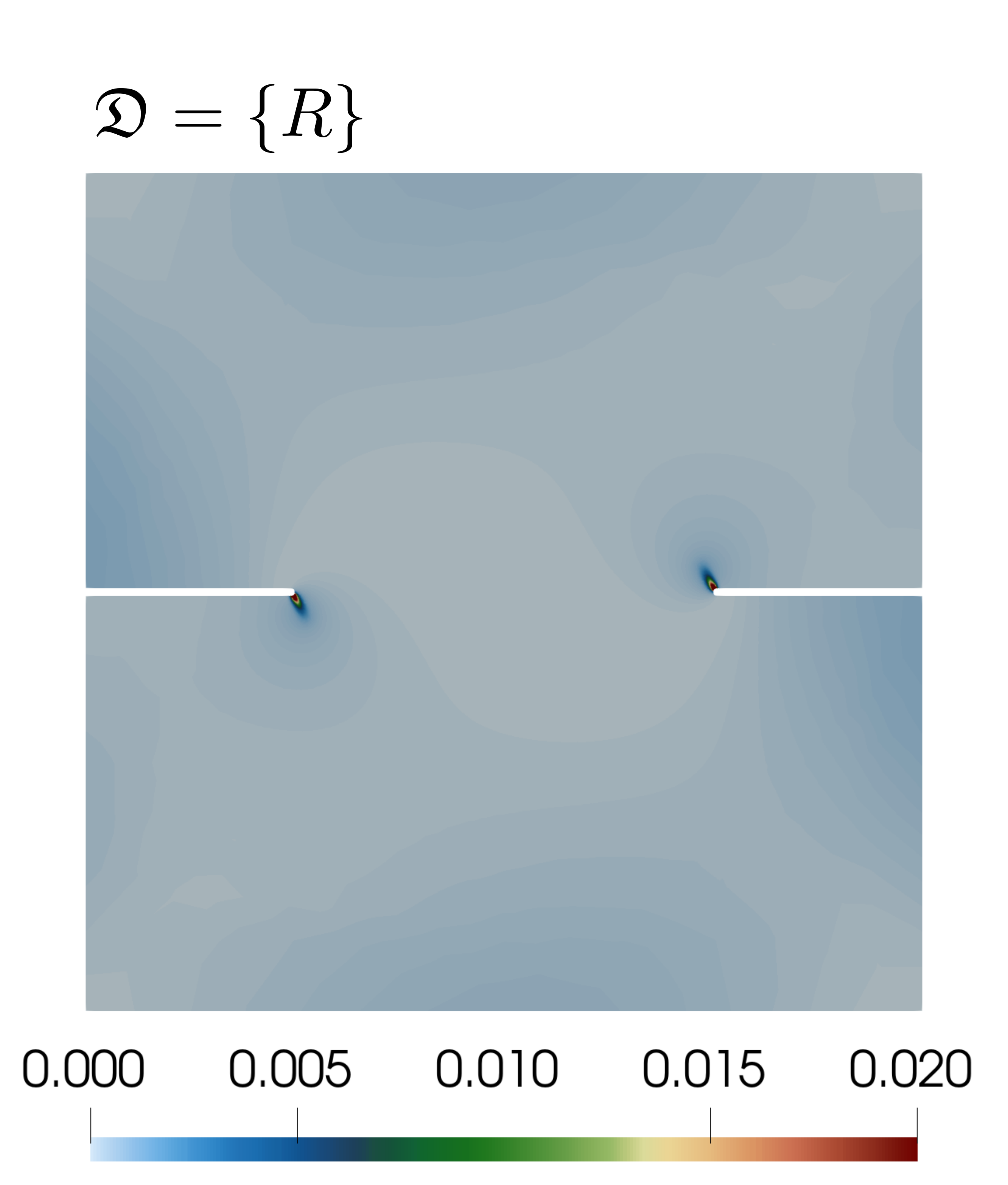}}
\caption{Fictitious undegraded energy density part $\psi_e^i$ ($i \in \mathfrak{D}$), where $\bar{u}_1 = \bar{u}_2 = 0.05$ mm.}
\label{fig:doublenotch_individual_psi}
\end{figure}

Regarding partial degradation, recall Eq.~\eqref{eq:distinct_degradation} that energy density parts that corresponds to the set $\mathfrak{U}$ remains completely undamaged, such that $g_i(d) = 1$ for $i \in \mathfrak{U}$, where $\mathfrak{D} \cup \mathfrak{U} = \left\lbrace B, C, R \right\rbrace$ and $\mathfrak{D} \cap \mathfrak{U} = \emptyset$. 
Within this platform, we first explore the effects of each individual energy density part by considering three different settings: (a) $\mathfrak{D} = \left\lbrace B \right\rbrace$; (b) $\mathfrak{D} = \left\lbrace C \right\rbrace$; and (c) $\mathfrak{D} = \left\lbrace R \right\rbrace$. 
Fig.~\ref{fig:doublenotch_individual} shows the fracture patterns for double edge notched tests for three aforementioned cases where $\bar{u}_1 = \bar{u}_2 = 0.05$ mm. 
Under the same threshold energy density $\psi_\text{crit}$, partial Boltzmann degradation case shown in Fig.~\ref{fig:doublenotch_B} undergoes crack propagation, whereas degrading $\psi_e^C$ [Fig.~\ref{fig:doublenotch_C}] or $\psi_e^R$ [Fig.~\ref{fig:doublenotch_R}] exhibit small amount of damage accumulation at the crack tip, without complete rupturing. 
It reveals that in either force or displacement-driven setup, most of the elastic energy is stored in non-polar constituent; the amount of stored energy density parts: $\psi_e^B > \psi_e^C > \psi_e^R$. 
As illustrated in Fig.~\ref{fig:doublenotch_individual_psi}, only the fictitious undegraded Boltzmann energy density part locally exceeds the prescribed threshold $\psi_{\text{crit}} = 1.0$ kJ/m$^3$ through the cracks when $\mathfrak{D} = \left\lbrace B \right\rbrace$. 
Even though $\psi_e^C$ and $\psi_e^R$ do not exceed the threshold energy except the flaw tip region, the results also indirectly evidence that the coupling and pure micro-rotational energy density parts affect crack kinking, while the pure Boltzmann part mainly drives the crack to grow.

\begin{figure}[h]
\centering
\includegraphics[width=1.0\textwidth]{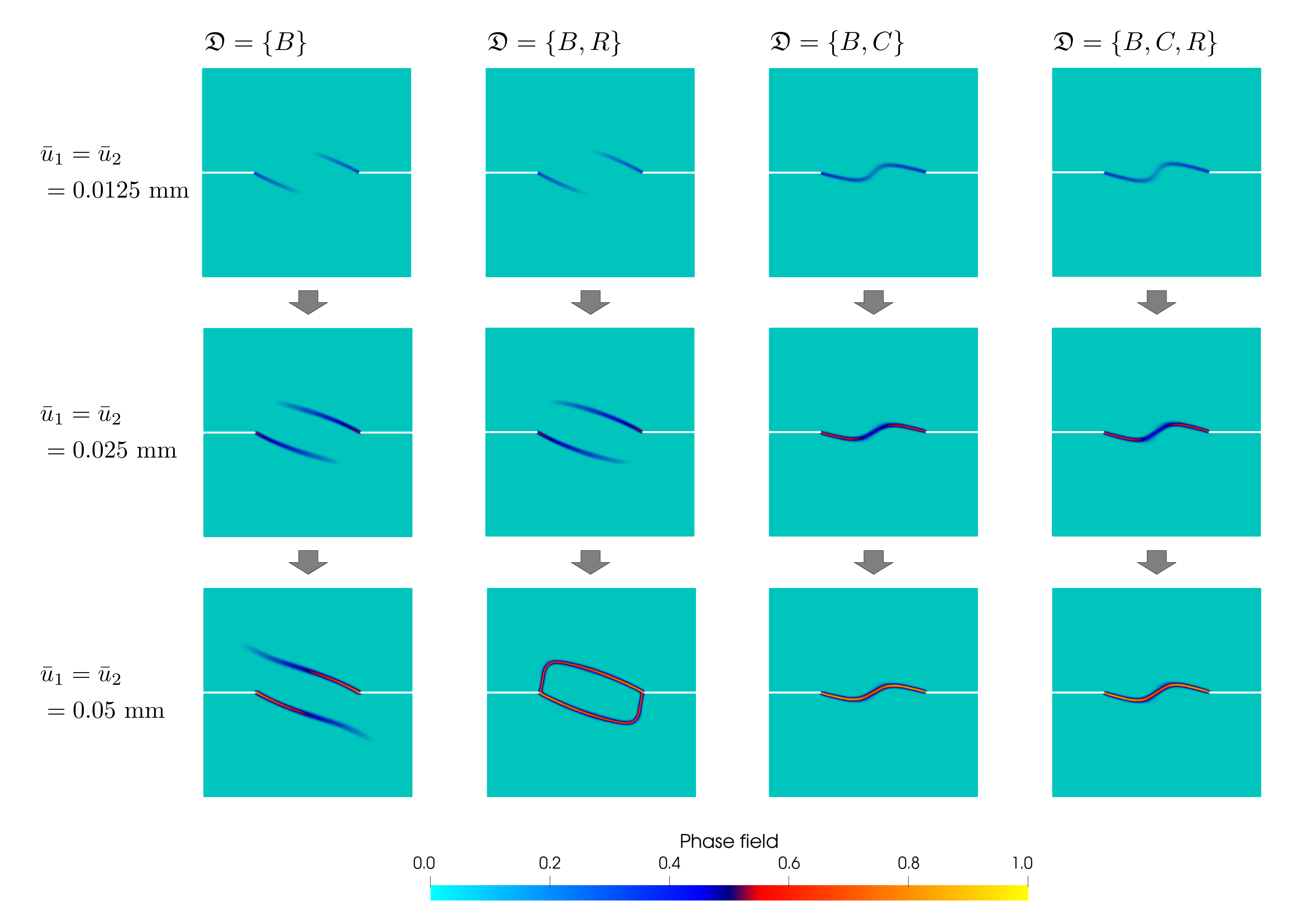}
\caption{Crack patterns for double edge notched tests with different combination of degrading energy density parts at several load increments.}
\label{fig:doublenotch_crackpath}
\end{figure}

\begin{figure}[h]
\centering
\includegraphics[height=0.375\textwidth]{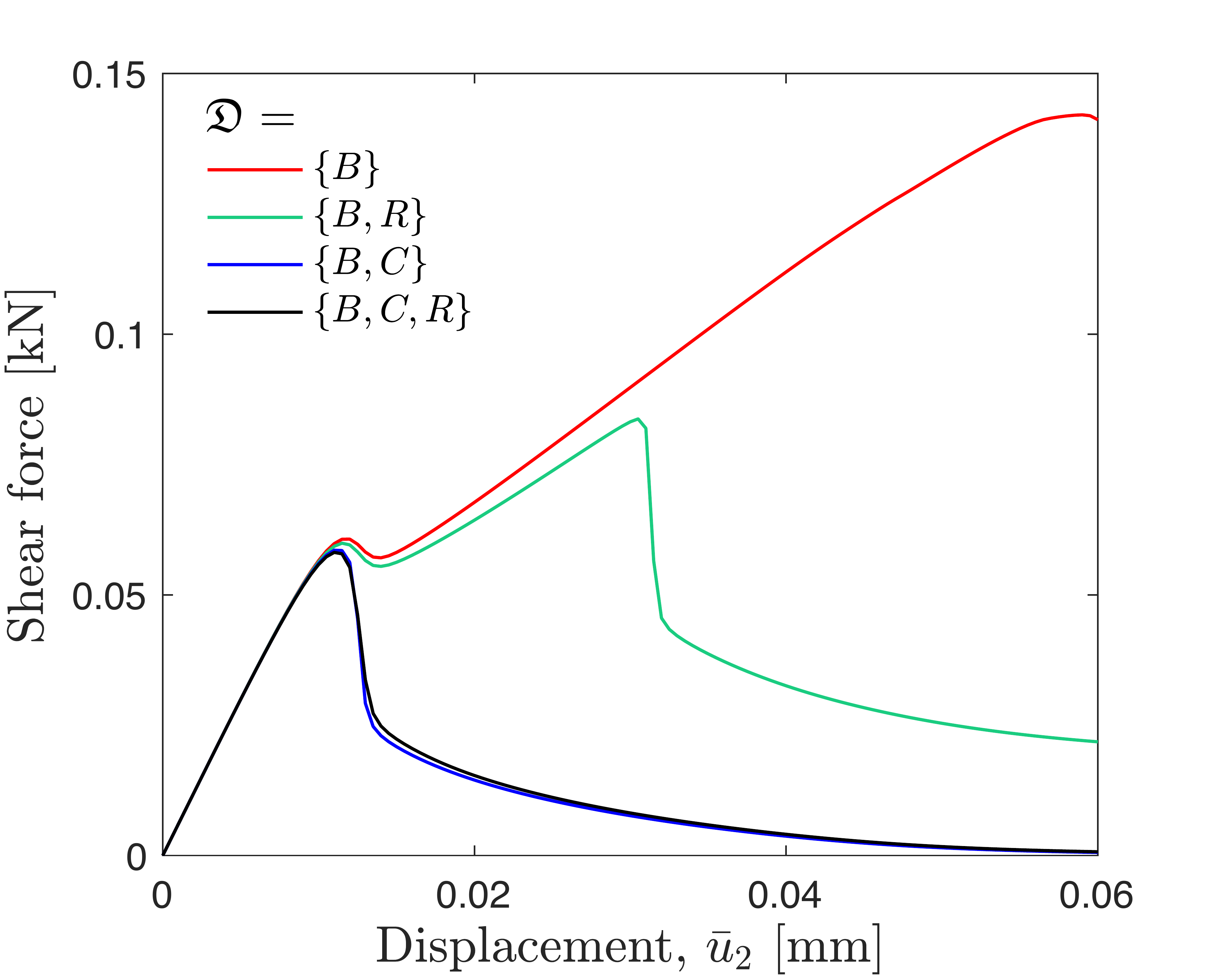}
\caption{The force-displacement curves from the double edge notched tests with different combination of degrading energy density parts.}
\label{fig:doublenotch_response}
\end{figure}

Since the pure Boltzmann part mainly drives the crack propagation, we now focus on the combined partial degradation with $B \in \mathfrak{D}$, also by considering three different settings within the same platform: (a) $\mathfrak{D} = \left\lbrace B,R \right\rbrace$; (b) $\mathfrak{D} = \left\lbrace B,C \right\rbrace$; and (c) $\mathfrak{D} = \left\lbrace B,C,R \right\rbrace$. 
Fig.~\ref{fig:doublenotch_crackpath} shows the crack patterns for double edge notched tests for three different combinations of partial degradation compared with the case where $\mathfrak{D} = \left\lbrace B \right\rbrace$, while Fig.~\ref{fig:doublenotch_response} illustrates the obtained load-deflection curves. 
The results confirms that degradation of the energy density parts $\psi_e^C$ and $\psi_e^R$ affects the crack kinking and curving. 

The combined degradation with $\mathfrak{D} = \left\lbrace B,R \right\rbrace$ tend to stimulate similar fracture patterns compared to the partial Boltzmann degradation case until $\bar{u}_1 = \bar{u}_2 = ~0.04$ mm, and then the cracks start to propagate towards the notches. 
Revisiting Fig.~\ref{fig:doublenotch_individual_psi}, this again indicates that the crack trajectories tend to follow the path that maximizes the energy dissipation (i.e., crack growth towards the adjacent flaw when the stored $\psi_e^R$ at the tip becomes high enough). 
Similarly, the combined degradation with $\mathfrak{D} = \left\lbrace B,C \right\rbrace$ leads the cracks to grow towards the adjacent tip. 
In this case, however, the cracks tend to kink towards the adjacent notch from the beginning, and then two cracks coalescence toward each other after sufficient loading. 
Since the amount of stored coupling energy $\psi_e^C$ is greater than the pure micro-rotational part $\psi_e^R$, we speculate that the coupling energy part influences the crack pattern more significantly compared to the micro-rotational energy part, so that we thus observe similar fracture pattern when $\mathfrak{D} = \left\lbrace B,C,R \right\rbrace$. 
In summary, this numerical experiment highlight that the pure Boltzmann energy density drives the crack growth, while the micro-continuum coupling energy density mainly influences the kinking direction.

\section{Conclusion}
This study presents a phase field fracture framework to model cohesive fracture in micropolar continua. 
To the best of the authors' knowledge, this is the first ever mathematics model that employs the phase field fracture framework to simulate crack growth in materials that exhibit size effect in both elastic and damaged regimes. 
To replicate a consistent size effect for both the elastic deformation and crack growth mechanisms, we introduce a method to incorporate distinctive degradation mechanisms via an energy-split approach for the non-polar, coupling and micropolar energies, while adopting the pair of degradation and regularization profiles that enables us to suppress the sensitivity of the length scale parameter that regularizes the phase field. 
One-dimensional analysis and numerical experiments demonstrate that the quasi-quadratic degradation function combined with linear local dissipation function successfully suppress the sensitivity of the length scale parameter for phase field while successfully incorporate the size effect with a length scale parameter that can be measured via standard inverse problems for micropolar materials. 
This result is significant, as the insensitivity of the length scale parameter will allow one to use coarser mesh to run simulations for a scale relevant to field applications (e.g. geological formations, structural components), while still able to replicating the size effect exhibited by materials of internal structures.

\section{Acknowledgments}
The authors would like to thank two anonymous reviewers who have provided helpful suggestions and feedback that improved the manuscript. 
The first and corresponding authors are supported by the Earth Materials and Processes
program from the US Army Research Office under grant contract 
W911NF-18-2-0306. 
The corresponding author is supported by the NSF CAREER grant from Mechanics of Materials and Structures program
at National Science Foundation under grant contract CMMI-1846875, the Dynamic Materials 
and Interactions Program from the Air Force Office of Scientific 
Research under grant contracts FA9550-17-1-0169 and FA9550-19-1-0318.
These supports are gratefully acknowledged. 
The views and conclusions contained in this document are those of the authors, 
and should not be interpreted as representing the official policies, either expressed or implied, 
of the sponsors, including the Army Research Laboratory or the U.S. Government. 
The U.S. Government is authorized to reproduce and distribute reprints for 
Government purposes notwithstanding any copyright notation herein.

\bibliographystyle{plainnat}
\bibliography{main}

\end{document}